\documentclass[manuscript,screen,review=False]{acmart}

\AtBeginDocument{%
  }

\setcopyright{acmcopyright}
\copyrightyear{2018}
\acmYear{2018}
\acmDOI{XXXXXXX.XXXXXXX}

\acmConference[Conference acronym 'XX]{Make sure to enter the correct
  conference title from your rights confirmation emai}{June 03--05,
  2018}{Woodstock, NY}
\acmPrice{15.00}
\acmISBN{978-1-4503-XXXX-X/18/06}

\usepackage{xcolor}
\usepackage{subcaption}
\usepackage{graphicx}
\usepackage{caption}
\usepackage{lipsum}
\usepackage{amsmath,amsfonts}
\usepackage{algorithmic}
\usepackage{siunitx}
\usepackage{textcomp}
\usepackage{rotating,multirow}
\usepackage{color,soul}
\definecolor{jmc}{RGB}{199, 0, 57}

\newcommand{\toolname}{SAT-MapIt}  
\newcommand{\technology}{TSMC \SI{65}{\nano\meter} LP}

\newcommand{\kreve}{\textit{reversebits}}

\newcommand{\ksqrt}{\textit{sqrt}}
\newcommand{\kstrs}{\textit{stringsearch}}
\newcommand{\kgsm}{\textit{gsm}}

\newcommand{\cmnt}[1]{}

\usepackage{threeparttable}
\usepackage{tabularx}
\usepackage{booktabs}
\usepackage{color, colortbl}

\usepackage{array}

\usepackage[acronym, toc,numberedsection = nolabel, section = part]{glossaries}
\makenoidxglossaries
\newacronym{alap}{ALAP}{As-Late-As-Possible}
\newacronym{asap}{ASAP}{As-Soon-As-Possible}
\newacronym{alu}{ALU}{Arithmetic-Logic Unit}
\newacronym{asic}{ASIC}{Application Specific Integrated Circuit}
\newacronym{cgra}{CGRA}{Coarse-Grain Reconfigurable Array}
\newacronym{cnf}{CNF}{Conjunctive Normal Form}
\newacronym{cil}{CIL}{Compute-Intensive Loop}
\newacronym{dfg}{DFG}{Data Flow Graph}
\newacronym{dma}{DMA}{Direct Memory Access}
\newacronym{fpga}{FPGA}{Field Programmable Gate Array}
\newacronym{ims}{IMS}{Iterative Modulo Scheduling}
\newacronym{isa}{ISA}{Instruction Set Architecture}
\newacronym{ilp}{ILP}{Integer Linear Programming}
\newacronym{ir}{IR}{Intermediate Representation} 
\newacronym{ii}{II}{Iteration Interval}
\newacronym{kms}{KMS}{Kernel Mobility Schedule}
\newacronym{mii}{mII}{minimum II}
\newacronym{ms}{MS}{Mobility Schedule}
\newacronym{pe}{PE}{Processing Element}
\newacronym{ra}{RA}{Register Allocation}
\newacronym{rtl}{RTL}{Register Transfer Level}
\newacronym{sat}{SAT}{satisfiability}
\newacronym{soa}{SoA}{State of the Art}
\newacronym{u}{U}{Utilization}

\glsdisablehyper
\renewcommand*{\a}[1]{\gls{#1}}
\newcommand*{\as}[1]{\glspl{#1}}

\newcommand{\secref}[1]{\hyperref[{#1}]{Section \ref*{#1}}}
\usepackage{enumitem}

\begin{document}


\title{SAT-based Exact Modulo Scheduling Mapping for Resource-Constrained CGRAs}

\author{Cristian Tirelli}
\email{cristian.tirelli@usi.ch}
\affiliation{%
  \institution{SYS Institute, Università della Svizzera Italiana}
  \city{Lugano}
  \country{Switzerland}
}

\author{Juan Sapriza}
\author{Rubén Rodríguez Álvarez}
\affiliation{%
  \institution{EPFL}
  \city{Lausanne}
  \country{Switzerland}}

\author{Lorenzo Ferretti} 
\affiliation{%
  \institution{Micron Technology}
  \city{San Jose}
  \country{United States}
}

\author{Benoît Denkinger}
\author{Giovanni Ansaloni}
\author{José Miranda Calero}
\author{David Atienza}
\affiliation{%
  \institution{EPFL}
  \city{Lausanne}
  \country{Switzerland}}

\author{Laura Pozzi} 
\email{laura.pozzi@usi.ch}
\affiliation{%
  \institution{SYS Institute, Università della Svizzera Italiana}
  \city{Lugano}
  \country{Switzerland}
}


\renewcommand{\shortauthors}{Tirelli et al.}

\begin{abstract}

Coarse-Grain Reconfigurable Arrays (CGRAs) represent emerging low-power architectures designed to accelerate Compute-Intensive Loops (CILs). The effectiveness of CGRAs in providing acceleration relies on the quality of mapping: how efficiently the CIL is compiled onto the platform. State of the Art (SoA) compilation techniques utilize modulo scheduling to minimize the Iteration Interval (II) and use graph algorithms like Max-Clique Enumeration to address mapping challenges. Our work approaches the mapping problem through a satisfiability (SAT) formulation. We introduce the Kernel Mobility Schedule (KMS), an ad-hoc schedule used with the Data Flow Graph and CGRA architectural information to generate Boolean statements that, when satisfied, yield a valid mapping. Experimental results demonstrate SAT-MapIt outperforming SoA alternatives in almost 50\% of explored benchmarks. Additionally, we evaluated the mapping results in a synthesizable CGRA design and emphasized the run-time metrics trends, i.e. energy efficiency and latency, across different CILs and CGRA sizes. We show that a hardware-agnostic analysis performed on compiler-level metrics can optimally prune the architectural design space, while still retaining Pareto-optimal configurations. Moreover, by exploring how implementation details impact cost and performance on real hardware, we highlight the importance of holistic software-to-hardware mapping flows, as the one presented herein.
\end{abstract}



\received{XX XXX 2023}
\received[revised]{XX March 2023}
\received[accepted]{XX June 2023}

\maketitle

\glsresetall

\section{Introduction}
\label{sec:intro}

The constant growth of computational requirements in everyday applications has increased the demand for high-performance and low-power architectures. Such architectures need to perform compute-intensive tasks efficiently, while at the same time dealing with tight power/resource constraints.

While \as{asic} accelerators have been largely adopted in these scenarios due to their efficiency, they are limited by their fixed functionality and their lengthy and costly design time. The reconfigurability of \as{fpga} can address these limitations, but at the cost of lower power and area efficiency, because, due to their fine-grained structure, the hardware overhead required to reconfigure  an \a{fpga} can add power and area overheads of up to $10\times$ and $20\times$, respectively~\cite{lin2014finegrain,kuon2006measuring}. 



\begin{figure}[t!]
\centering
  \includegraphics[width=\textwidth]{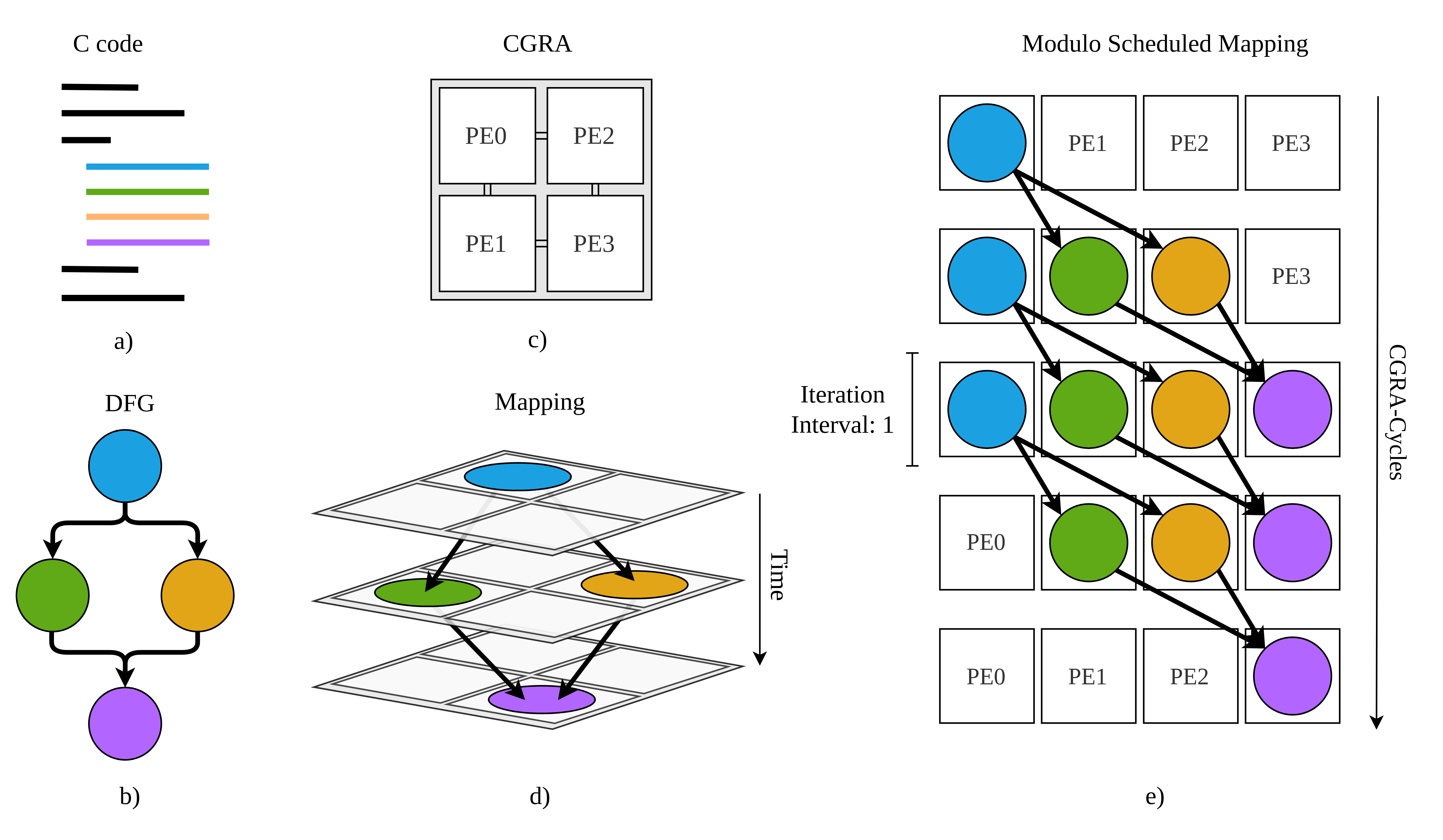}
  
    \caption{Overview of the \glsentryshort{cgra} scheduling and mapping workflow. a): A \glsentryshort{cil} is identified in the source C code, b): Its associated \glsentryshort{dfg} is derived by compiler analysis. c): An abstract view of available hardware resources  (i.e., number of processing elements and their connectivity) is constructed. d): The nodes of the \glsentryshort{dfg} are mapped on the processing elements while complying with architectural constraints. e): Nodes are modulo scheduled across different \glsentryshort{cgra}-instructions, partially overlapping the execution of multiple iterations. In the example, a new iteration is started at each \glsentryshort{cgra}-instruction, hence the scheduling has an Initiation Interval (II) equal to 1. }
  \label{fig:workflow-overview}
\end{figure}

\as{cgra} provide a meet-in-the-middle approach, providing run-time reconfigurability at the level of arithmetic operations, with very high computational efficiency \cite{li2021chordmap, karunaratne2017hycube}. These characteristics are well suited in domains such as streaming and multimedia applications \cite{li2021chordmap, akbari2018px, karunaratne2017hycube, oh2009recurrence, lee2015optimizing, wijerathne2019cascade, duch2017heal}, but also in the edge domain, where they have been shown to enable flexible hardware acceleration in resource-constrained scenarios. 


A \a{cgra} is a mesh of \as{pe} organized in a two-dimensional grid; each \a{pe} contains an \a{alu} and a number of internal registers.  \as{pe} are connected to their neighbors according to a mesh topology. In addition, shared connections are usually present from \as{pe} to external memory to load inputs and store outputs. Computation in \as{cgra} is organized in \a{cgra}-cycles, during which every \a{pe} performs an operation. The list of supported operations defines the \a{isa} of a \a{cgra}.
Therefore, in addition to distributing operations spatially throughout its \as{pe}, most \as{cgra} can execute different instructions at each clock cycle, allowing to map computational kernels requiring more operations than the amount of available \as{pe}. Time-multiplexed CGRAs offer high task flexibility and hardware reuse, enabling the acceleration of a wide set of tasks at a minimal area and power overhead. In the case of~\cite{RodriguezMay2023}, the area of the \a{cgra} is dominated by the datapath and the power overhead of configuring \as{pe} at  every cycle is only $\sim50\%$. This comes in stark contrast with the reconfiguration required by FPGAs, which affects large portions of the fabric and demands a hardware overhead that reflects on $~10\times$ dynamic power increase with respect to an ASIC implementation~\cite{lin2014finegrain,kuon2006measuring}. Although  CGRAs that only support spatial mapping exists (\cite{govindaraju2012dyser, wu2014q100, prabhakar2017plasticine,renesasstp}),  their lack of flexibility in the time dimension leads to much higher number of \as{pe}.
especially when control operations have to be performed ~\cite{karunaratne20194d,nguyen2021fifer}.

While large CGRAs meshes can be the preferred choice in high-performance architectures, edge
scenarios have to cope with tight power and area budgets. In this cases, small, time-multiplexed
CGRAs constitute a promising avenue towards the design of versatile and energy-efficient accelerators. Indeed, important energy and latency savings with respect to low-power CPU execution have been reported in the \a{soa} from employing resource-constrained \as{cgra}. In ~\cite{Duch:227874} a 37\% energy decrease is reported when running compressed-sensing algorithms. \cite{Carpentieri2024} showcases $3.4\times$ and $9.9\times$ energy and latency improvements, respectively, when performing convolutions. The architecture of~\cite{kim2014ULPSRP} reports energy improvements of almost 60\% when compared to a CPU-only platform from the \a{soa}.

One of the fundamental challenges of \a{cgra} exploitation is the compilation process, i.e. the translation of high-level source code onto \a{cgra} code, while taking advantage of the parallelism offered by the architecture. To do so, a technique called modulo scheduling is traditionally employed, which constructs a \a{dfg} from a \a{cil}, and then maps \a{dfg} nodes onto architecture \as{pe} in an interleaved manner so that nodes from different iterations coexists in the same \a{cgra}-cycle, minimizing the overall latency of each iteration.

In this work we target the challenge of mapping \as{cil} into resource constrained \as{cgra} by employing a modulo scheduling approach. To this end, we herein target the open-hardware OpenEdgeCGRA architecture of \cite{RodriguezMay2023} to demonstrate how our approach enables an effective design-space exploration to better exploit the \a{cgra}'s capabilities. The targeted \SI{65}{\nano\meter} technology implementation requires an area as little as \SI{0.12}{\milli\meter\squared} (for a minimalistic $2\times2$ mesh), and can offer up to $3.8\times$ and $3.5\times$ speed-up and energy gains respectively when compared to a low power CPU in control-instensive loops.

\autoref{fig:workflow-overview} portrays an overview of this process, where a section of C code is extracted and mapped into a $2\times2$ \a{cgra}. First, a loop is marked for \a{cgra} acceleration. Then, its \a{dfg}, which illustrates operations as nodes and data dependencies as edges, is derived.  The \a{dfg} is then scheduled over the \as{pe} of the target architecture, at different \a{cgra}-cycles, while abiding to resource and dependency constraints. A possible modulo scheduling of the simple \a{dfg} in \autoref{fig:workflow-overview}.b is shown in \autoref{fig:workflow-overview}.e, highlighting the interleaving among loop iterations.



As the number of \as{pe} (and the resources available to them) is reduced to fit on edge devices, mapping tools are faced with growing constraints. Existing techniques for modulo scheduling, described in \secref{sec:related_works}, rely mainly on heuristics to schedule, place, and then route operations and data on the \as{pe}. 
Here, we propose to address the mapping problem using \toolname{}, a \a{sat}-based formulation where data dependency, architectural constraints, and schedule are expressed as Boolean constraints. A valid mapping is then identified by determining if an assignment of the variables exists to satisfy the constructed Boolean formula.
We show that this technique is able to efficiently explore the space of possible mappings better than \a{soa} techniques, hence producing high-performance mappings even for a tightly constrained \a{cgra} topology. 

Building upon~\cite{Tirelli2023} we further clarify the toolchain's architecture, enriching the SAT formulation and the description of the register allocation phase to ensure a deeper understanding of its functionality and efficiency. 
Additionally, we showcase the performance of our approach as part of a complete code-to-hardware framework, allowing the evaluation of speedups with respect to non-accelerated executions and contrasting  mapping figures of merits with run-time metrics.
The flow starts from annotated C code and has at its endpoint a configuration instructions list (\textit{bitstream}) that governs the execution of a (simulated or synthesized) \a{cgra}. To this end, we herein target the open-hardware \a{cgra} design in \cite{RodriguezMay2023}. We highlight that compiler-level metrics (such as \a{ii} and \a{u}) do indeed correlate with the run-time performance of a mapped \a{cil}. However, hardware analysis (and hence the availability of a link between the compiler and the hardware) is the key to accurately assess the energy and timing requirements for a given \a{cil} and \a{cgra} architecture.

This paper is organized as follows. \secref{sec:related_works} reviews existing literature; 
\secref{sec:background} introduces foundational elements of our work and our problem formulation; \secref{sec:methodology} describes our \a{sat}-based methodology and \secref{sec:results} compares the results obtained by \toolname{} with respect to \a{soa} alternatives. \secref{sec:arch} illustrates the developed code-to-hardware framework used to validate our approach targeting resource-constrained hardware while \secref{sec:results_hardware} shows execution measurements and analyzes the mapping and run-time performance metrics to prune the design space. Finally, \secref{sec:conclusion} concludes the article.

\section{Related Work}\label{sec:related_works}

A recent survey \cite{podobas2020survey} has summarized the evolution of \a{cgra} architectures and methodology advances in the last thirty years of research. Herein, we  focus in particular on the \as{cgra} mapping problem, i.e. the compilation of software source code onto \a{cgra} code, defined as the mapping of \as{cil} instructions onto \as{pe}.
Existing methodologies can be divided into two main categories: the first using heuristics and the second using exact solutions. 

Early works in the first category includes the method proposed by Mei et al.\cite{mei2007adres}, which formulates scheduling, placement, and routing problems altogether, and proposes to solve them using simulated annealing. The attempt to solve all problems jointly does lead to long execution times, and potentially also to low quality mapping due to the difficulty to find good solutions in such a large space. 
Alternatively, in \cite{park2008edge} an edge-centric approach to modulo-scheduling was presented, where a schedule is generated by first routing each edge in the dataflow graph, and placement is addressed as a by-product of routing. In \cite{hamzeh2012epimap}, EPImap proposed to use both routing and re-computation to find a valid mapping, by adopting an epimorphic (time-extended) graph formulation enabling efficient identification of valid solutions. By adding nodes to the data dependency graphs, EPImap can find new valid solutions to the mapping while re-scheduling is performed to improve the efficiency.
EPImap's performance was later improved by GraphMinor \cite{chen2014graph} and REGIMap \cite{hamzeh2013regimap},  by reducing the mapping problem to the graph minor and max clique problem. In turn, \cite{dave2018ramp} RAMP authors have further refined REGIMap by  explicitly modeling and exploring various routing strategies and choosing the best one for each given \a{cil}. CRIMSON\cite{balasubramanian2020crimson} then proposed a randomized \a{ims} algorithm that explored the scheduling space more efficiently, and PathSeeker \cite{pathseeker} improved on CRIMSON~\cite{balasubramanian2020crimson} by analyzing mapping failures and performing local adjustments to the schedule to obtain a shorter compilation time and better quality of the solution.


In our experiments, we compare our results to those obtained by both RAMP\cite{dave2018ramp} and PathSeeker\cite{pathseeker}. These two works had shown superior performance with respect to the earlier methodologies mentioned above, and therefore represent the current \a{soa} of the modulo scheduling mapping problem. We quantitatively compare our work to them, and show that \toolname\ can better explore the scheduling space and get smaller \a{ii} by using custom scheduling tables with a \a{sat} formulation.

A second category of approaches addresses the mapping problem  with \a{ilp} or Boolean Satisfiability formulations. In \cite{chin2018architecture} the authors propose an \a{ilp} formulation approach and prove the feasibility of mapping in the given number of \a{cgra}-cycles. Similarly, \cite{miyasaka2020sat} propose to use a \a{sat} solver instead of an \a{ilp} solver to identify a valid solution. 

The SAT
formulation proposed in this paper 
addresses a critical limitation of the approach in [22], by managing back-edges in dataflow graphs and hence being capable of modelling
loop-carried dependencies, where the formulation in [22] could only consider a feed-forward, pipelined solution. This is an important advancement both theoretical -- because in the absence of back-edges the problem admits a more trivial solution -- and practical, as the capability to process loop-carried dependencies results in the ability to accelerate more diverse loops.

Our \a{sat} formulation, initially introduced in~\cite{Tirelli2023}, is to the best of our knowledge the first SAT-based solution to the modulo scheduling problem.

Most \a{soa} work lacks a direct link to the execution of mapped \as{cil} on hardware. In fact, \cite{hamzeh2012epimap, chen2014graph, hamzeh2013regimap, oh2009recurrence, park2008edge} all adopt abstract representations as architectural targets for their scheduling frameworks. This approach may lead to simplifying assumptions that do not reflect on hardware. As examples, \cite{chen2014graph} and \cite{hamzeh2012epimap} postulate that there are no constraints on instruction memory, while in \cite{hamzeh2012epimap} and \cite{hamzeh2013regimap} a large number of concurrent data memory transfers are allowed, without considering the capability of the system bus. An alternative approach is followed by the authors of \cite{dave2018ramp} and \cite{wijerathne2022morpher}, who embed their \a{cgra} compilation methodologies as part of system simulation frameworks, based on gem5 \cite{gem5} and SystemC, respectively. In this way, they ensure that the defined target hardware  can be properly integrated in a system-on-chip. Nevertheless, system simulation waives the detailed description of the hardware implementation and hence does not offer a path to digital verification and synthesis, hampering the insights which can be gathered from an architectural perspective. Conversely, this works builds upon~\cite{Tirelli2023} to interface to an open-hardware \a{cgra} implementation \cite{RodriguezMay2023}. In this way, we ensure the cycle-accurate correctness of \toolname{} mapping via post-synthesis simulations, reporting the performance and requirements of different \a{cgra} sizes. Other works adopting this position are \cite{pathseeker} and \cite{chin2018architecture}, where scheduled \as{cil} are mapped on the CGRA-ME array \cite{chin2017cgra}. Of these, PathSeeker \cite{pathseeker} is  most related to our work, since it also implements modulo scheduling. We comparatively evaluate our \toolname{} methodology against it in \secref{sec:results}.

\begin{figure*}[t]
\centering
\minipage[t][1cm][t]{0.24\textwidth}
  \includegraphics[width=\linewidth]{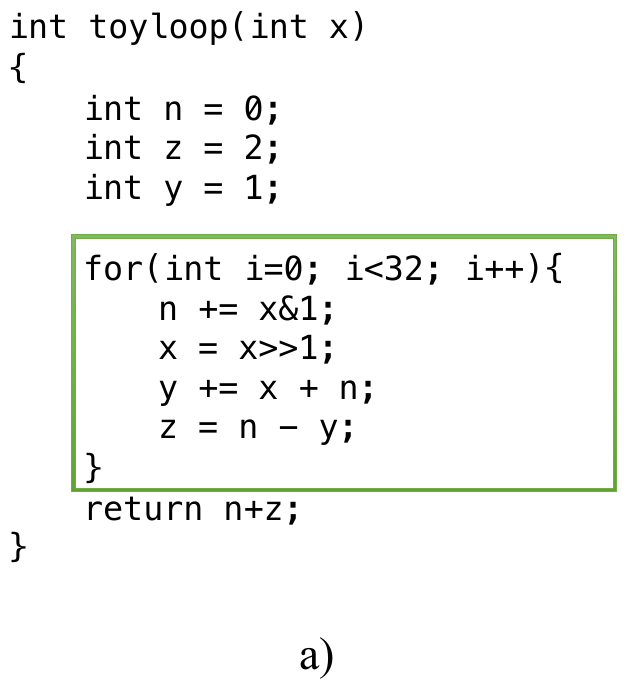}
\endminipage\hfill
\minipage[t][1cm][t]{0.34\textwidth}
  \includegraphics[width=\linewidth]{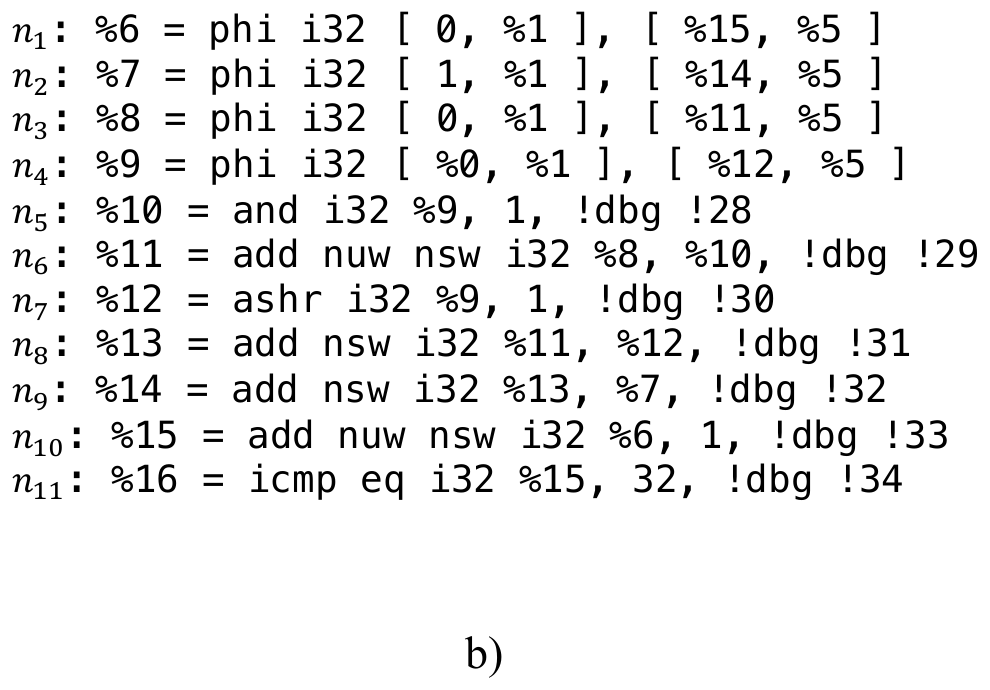}
\endminipage\hfill
\minipage[t][2cm][t]{0.3\textwidth}
  \includegraphics[width=0.6\linewidth]{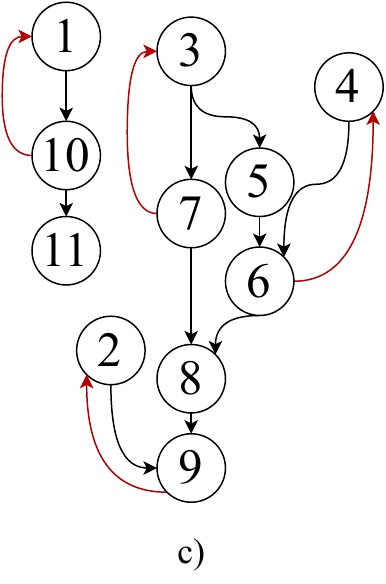}
\endminipage\hfill
\caption{a): C code of our running example. The identified \a{cil} is highlighted.  b): LLVM \glsentryshort{ir} of the identified \a{cil}. c): Associated \a{dfg}. \\Red edges are loop-carried dependencies, black edges are data dependencies.}
\label{fig:codetodfg}
\end{figure*}

\section{Background}
\label{sec:background}

In this section, we provide the background needed to present our methodology, and we illustrate it wherever appropriate through a running example.

\subsection{Compilation}

To accelerate an application to a \a{cgra}, a \a{cil} is identified in the application, as in \autoref{fig:codetodfg}.a; the identification can either be performed automatically through techniques such as, for example, \cite{zacharopoulos2018regionseeker}, or manually by the programmer through pragma annotations, as done in this work. Then, \a{cil} must be compiled, in order to be translated into \a{cgra} instructions. 

Compilation is in general the process of translating a program written in high-level code into a semantically-equivalent one in low-level code. The first step in this process, depicted in \autoref{fig:codetodfg}.b is to generate a  version in \a{ir} (LLVM \a{ir} is our chosen one) from the original high-level-code. \\The second step is to go from there to \a{dfg}: \as{dfg} are directed graphs in which nodes represent operations, edges represent dependency relations between operations, and loop-carried dependencies correspond to back-edges. Note that the \a{dfg} reflects the number and types of instructions in the LLVM IR, which may differ from those seen in the original high-level code, as an effect of semantic-preserving compiler transformations.

In a second phase, the generated \a{dfg} is mapped onto the \a{cgra}, by assigning each of its nodes to a given \a{pe} in a given \a{cgra}-cycle. To perform such mapping, a technique called modulo scheduling is employed and is described in the following.

%

\subsection{Modulo Scheduling}

\begin{figure}[t!]
    
    \begin{subfigure}{0.3\textwidth}
    \includegraphics[width=\linewidth]{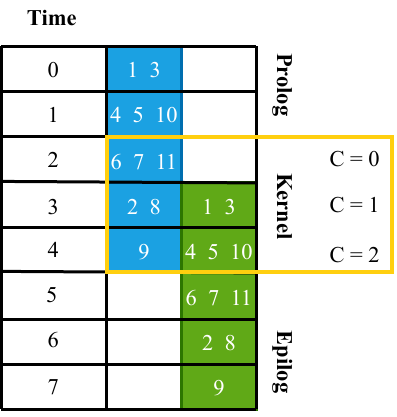}
    \caption{} \label{fig:mapresa}
  \end{subfigure}%
  \hspace{1cm}   
  \begin{subfigure}{0.41\textwidth}
    \includegraphics[width=\linewidth]{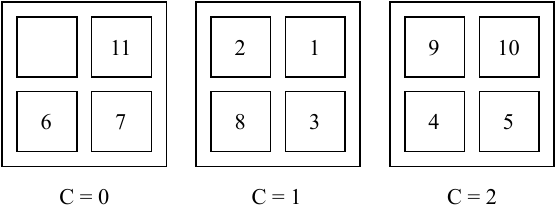}
    \caption{} \label{fig:mapresb}
  \end{subfigure}%

    \caption{a): Modulo scheduling of the \a{dfg} of the running example, highlighting the division between prologue, kernel, and epilogue. \\b): Mapped \a{cil} of the \a{dfg} in the running example on a $2\times2$ \a{cgra}}
    \label{fig:mapres}
\end{figure}

Modulo scheduling is a compilation technique that enables efficient execution of a \a{cil}, by executing multiple iterations of it in an interleaved manner. As exemplified in \autoref{fig:mapresa}, a modulo-scheduled \a{cil} is divided into three stages: prologue, kernel, and epilogue. Prologue and epilogue are one-time executed stages: the former is used to prepare the data to feed the pipeline, the latter to reorganize them at the end. The kernel is instead the steady state that executes multiple times and includes the operations to be parallelized through pipelining.
The goal of modulo scheduling is to  pipeline as effectively as possible the execution of \a{cil} operations, and this corresponds to minimizing the \a{ii}, i.e. the duration of the kernel stage -- 3 \a{cgra}-cycles in our example.
Therefore, the mapping problem consists in finding a legal modulo schedule for a \a{cil},  performing the placement and routing of operations in a constrained 3D space represented by the \as{pe} dimensions and by time. An example of legal mapping for the \a{dfg} in our running example is shown in \autoref{fig:mapres}.b.

\subsection{Problem Formulation}

The mapping problem requires taking into account architectural constraints such as the number of available \as{pe}, the available interconnections among \as{pe} and memory, and the number of registers for each \a{pe}. Violation of such constraints cause invalid mappings leading to, e.g., the schedule of more operations than the number of available \as{pe} during the same \a{cgra}-cycle, or the impossibility to route a value between parent and children nodes. We formally define the problem as following:

\textit{Mapping Problem}: Given a \a{dfg} $\mathcal{G(N,E)}$ representing a \a{cil}, and given a \a{cgra} architecture $\mathcal{A}$ in terms of size and topology, we generate a set of literals $\mathcal{L}$ in the form $v_{n,p,c,it}$  where $n$ is the node id, $p$ is the \a{pe} on which $n$ is mapped, $c$ is the \a{cgra}-cycle at which $n$ is scheduled and $it$ is the iteration to which $n$ refers.
We encode all constraints to which a mapping has to adhere to, in order to be valid, via a \a{sat} formulation.

\subsection{SAT problem}

A Boolean satisfiability problem, namely (\a{sat}), is the problem of determining if a Boolean formula is satisfiable or unsatisfiable. Usually, a problem is formulated in a \a{cnf}: a conjunction of clauses, each being a disjunction of literals.

For example, \autoref{satformulation} shows a \a{cnf} formula consisting of three literals $a$, $b$, and $c$, and three statements $a\vee b$, $b$, $c\wedge a$.

\begin{equation}\label{satformulation}
    (a\vee b)\wedge b \wedge (c\wedge a)
\end{equation}

In order to determine the satisfiability of the \a{cnf} formula in \autoref{satformulation} a \a{sat} solver searches for an assignment of the literals such that the formula is true. In this example, the answer is \a{sat}, because setting a, b and c to true makes the formula evaluate to true.

In our work we must impose that only a fixed number of literals can be true at the same time. This constraint is called an at-most-K constraint and is well known in the literature\cite{bittner2019sat}.

Encoding the mapping problem within the framework of a SAT problem offers several distinct benefits. 
The primary advantage of using SAT solvers in our methodology is to exploit their capability to conduct a more structured and comprehensive exploration of the solution space and, at the same time, find exact solutions. Ensuring high-quality mappings that are critical to achieving high performance in constrained scenarios. This is especially relevant for low- to moderate-sized CGRA configurations, where the solution space, while large, remains within the tractable bounds for exact methods.
A further benefit is the utilization of established solvers, such as Z3, which underscores the strategic integration of robust, industry-standard tools in our methodology.

In the following, we detail the \a{sat} formulation we devised to solve the mapping problem.


\section{CGRA mapping Methodology}
\label{sec:methodology}

Our methodology addresses mapping by solving a \a{sat} problem where data dependency, schedule and \a{cgra} architecture are expressed as Boolean constraints in a \a{cnf}.
Our toolchain takes as input the C code of the program, converts it into LLVM \a{ir}, extracts the designated \a{cil} structures and, through a custom LLVM pass, obtains the information required to generate the \a{dfg}.

\begin{figure}[t]
\centering
  \includegraphics[width=0.5\textwidth]{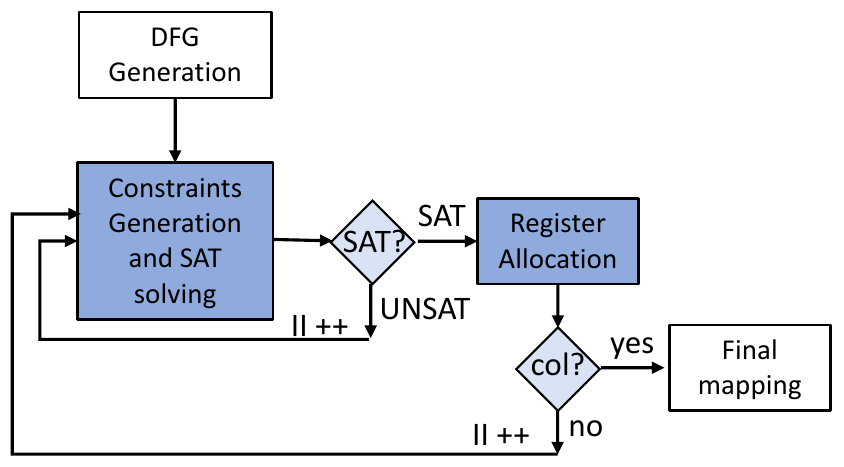}
  \caption{\toolname\ searches for mappings for a given \a{ii}, iteratively increasing \a{ii} in case the \a{sat} solver returns UNSAT, or register allocation fails to color the  model returned by the solver.}
  \label{fig:toolchain}
\end{figure}


The information retrieved from the LLVM pass
is then used to build a set of schedules, in turn translated into a set of constraints, which are finally fed to a \a{sat} solver, as  depicted in \autoref{fig:toolchain}. If the answer of the solver is \a{sat}, the next step is to verify that there are enough registers available in the \as{pe} to store the generated data for the whole liveness duration, and this is determined via \a{ra}. If this further step also succeeds, then a valid mapping has been identified. Otherwise, the current \a{ii} increases and the process is repeated iteratively.





Here, we detail the various steps of our methodology: schedule creation, \a{sat} formulation, and finally register allocation.

\subsection{Schedule creation}\label{subsec:schedule_creation}

We first create \a{asap} and \a{alap} schedules, for the input \a{dfg}. This corresponds to detecting how early and how late each node can be scheduled. 
We then generate the \a{ms}, which expresses the mobility of each node from its $ASAP$ to its $ALAP$ position.  \autoref{tab:scheds} shows these schedules for our running example. 

\begin{table}[h]
    \caption{ASAP, ALAP, and Mobility Schedule}
    \centering
    \begin{tabular}{c|lll}
        \hline
         &  \multicolumn{3}{c}{\textbf{Nodes}}\\ 
         \textbf{Time}&  \textbf{ASAP}&  \textbf{ALAP}& \textbf{MS}\\ \hline
         0&  1 2 3 4&  3& 1 2 3 4                          \\ 
         1&  5 7 10&  4 5& 1 2 4 5 7 10                     \\
         2&  6 11&  1 6 7& 1 2 6 7 10 11                     \\
         3&  8&  2 8 10& 2 8 10 11                   \\
         4&  9&  9 11& 9 11                             \\ \hline
    \end{tabular}
    \label{tab:scheds}
\end{table}

At this point, we create the \a{kms}: an ad hoc structure used to formulate the mapping problem with \toolname.
The \a{kms}  can be seen as \emph{a superset of all possible mappings}, and is the product of iteratively folding \a{ms} by an amount equal to $\a{ii}$: every time \a{ms} is folded by $\a{ii}$ into \a{kms}, each node receives a label that refers to the iteration number it belongs to. The more folding we have, the more iterations our \a{cil} will execute in the kernel. This process is depicted in \autoref{tab:kmsgen}, again for our running example. Given an $\a{ii}$ of 3, the \a{ms} is folded twice ($\lceil 5/3 \rceil = 2$) and, hence, the \a{kms} contains two iterations. The first is depicted in blue, and the second in green.

\definecolor{green}{rgb}{0.37,0.7,0.1}

\begin{table}[h]
\centering
    \caption{Kernel Mobility Schedule.}
        \begin{tabular}{c|ll}
         \multicolumn{2}{c}{\textbf{KMS}}\\
        \hline
         \textbf{Time}& \multicolumn{2}{c}{\textbf{Nodes}}\\
         \hline
         0& $1_0 2_0 6_0 7_0 10_0 11_0$ &  \\
         1& $2_0 8_0 10_0 11_0$ & $1_1 2_1 3_1 4_1$ \\
         2& $9_0 11_0$ & $1_1 2_1 4_1 5_1 7_1 10_1$\\
         \hline
         \multicolumn{3}{c}{}\\
         \multicolumn{3}{c}{}\\
    \end{tabular}
    \label{tab:kmsgen}
\end{table}

Together, \a{dfg} and MKS are used to generate all the statements of the \a{cnf} formulation of the mapping problem, which is the subject of the next subsection. The iterative search for a valid mapping can be shortened by considering that available resources and loop carried dependencies impose a lower bound on the achievable II. Such \a{mii} as defined in~\cite{rau1996iterative}, is hence expressed as

\begin{equation}
\a{mii} = max(ResII, RecII)
\label{eq:mii}
\end{equation}
The first term provides the lower bound derived from the resource usage requirements of the operations to be mapped, and is computed as $ResII = \big\lceil \frac{\#nodes}{\#\as{pe}}\big\rceil$. The second term
corresponds to the length of the longest loop $l$ of nodes  from one iteration to another across a loop carried dependency in the data flow graph, and is given by $RecII = max\left(\lceil\frac{length(l)}{distance(l)}\rceil\right)_{l \in DFG}$.  For the \a{dfg} in our running example, and a $2\times2$ \a{cgra}:  $ \big\lceil \frac{11}{2\cdot 2}  \big\rceil = 3$, while the longest loop of nodes from one iteration to the next one is 2. Hence, from \autoref{eq:mii}, $\a{mii} = max(3,2) = 3$, which means that no valid solution can exist for lower initiation interval given for this target application and architecture.

\subsection{SAT formulation}\label{subsec:sat_formulation}

We create a \a{cnf} formula using literals in the following form:
$x_{n, p, c, it}$
, where $n$ denotes the node identifier in the \a{dfg}, $p$ denotes a \a{pe} on the \a{cgra}, $c$ represents at which \a{cgra}-cycle a node is scheduled, and $it$ to  which iteration the node refers to. 
For example, the literal $x_{3,2,1,0}$ represents whether node 3 is mapped on PE2, at time 1 and iteration 0.

Our problem formulation can be described at a high level by partitioning all statements into three main sets of clauses that assure the following:
\begin{itemize}
\item C1: Every node is associated with a set of literals, and for each one of those sets, one and only literal must be set to True
\item C2: At most one node should be assigned to a \a{pe} at a given \a{cgra}-cycle, since two or more nodes cannot be scheduled simultaneously on the same \a{pe}.
\item C3: Each node's predecessor and/or successor must be assigned to a neighbor or on the same \a{pe}.
\end{itemize}
To provide a formal description of the above constraints, we introduce some additional definitions.
Let $\mathcal{L}$ be the set of all literals, then $\mathcal{L}(n)$ be the set of all literals associated with node $n$. For example, for node $n_3$ of the running example we would have: \begin{equation}
    \mathcal{L}(n_3)= \{x_{3,0,1,1},x_{3,1,1,1},x_{3,2,1,1},x_{3,3,1,1}\}
    \label{eq:esn3}
\end{equation}
because, in the \a{kms}, in \autoref{tab:kmsgen} node 3 appears only at time 1 and at iteration 1, and can be mapped in any \a{pe} 0,1,2,3. 

To make the notation more compact and easy to read, we also associate each literal in the form $x_{n,p,c,it}$ to a literal written as $v_i$.

So, \autoref{eq:esn3} can also be written as:
\begin{equation*}
    \mathcal{L}(n_3)= \{v_{0},v_1,v_2,v_3\}
\end{equation*}
where $v_0 = x_{3,0,1,1}$, $v_1 = x_{3,1,1,1}$, $v_2 = x_{3,2,1,1}$ and $v_3 = x_{3,3,1,1}$.

Now, we can start the description of the three sets of constraints. The first set of constraints, C1, ensures that all the nodes are mapped on the \a{cgra}, and can be encoded formally with:
\begin{equation}
	\begin{aligned}
		\phi(n) &= \bigvee_{v_i \in \mathcal{L}(n)} v_i \\
		\alpha(n) &=\bigwedge_{(v_i,v_j)\in \mathcal{M}(n)} \neg (v_i \wedge v_j)\\
		\beta(n) &= \phi(n) \wedge \alpha(n)\\ 
\label{eq:stm1}
	\end{aligned}
\end{equation}
where $n$ is one of the nodes in the \a{dfg} and $\mathcal{M}(n)$ is defined as follows:
\begin{equation*}
    \mathcal{M}(n) = \{(v_i,v_j): v_i \prec v_j, (v_i,v_j)\in \mathcal{L}(n) \times \mathcal{L}(n) \}
\end{equation*}

where $v_i = x_{n,p_1,c_1,it_1}$ and $v_j = x_{n,p_2,c_2,it_2}$ with $p_1 \neq p_2$, $c_1\neq c_2$ and $it_1\neq it_2$.
Furthermore the symbol $\prec$ represents the \textit{lexicographically smaller-than} relation between two literals.
For example $x_{3,0,1,0}$ is lexicographically smaller than $x_{3,1,0,0}$, while $x_{3,1,0,1}$ is not.

\autoref{eq:stm1} will be used on each node of the \a{dfg} and each $\beta$ generated will be added to the \a{sat} solver.
In our example with $n_3$ \autoref{eq:stm1} becomes:
\begin{equation*}
	\begin{aligned}
		\phi(n_3) &= v_0 \vee v_1 \vee v_2 \vee v_3 \\
		\alpha(n_3) &= \neg (v_0 \wedge v_1) \wedge \neg (v_0 \wedge v_2) \wedge \neg (v_0 \wedge v_3) \wedge\\
		& \neg (v_1 \wedge v_2) \wedge \neg (v_1 \wedge v_3) \wedge \neg (v_2 \wedge v_3)\\
		\beta(n_3) &= \phi(n_3) \wedge \alpha(n_3)\\ 
	\end{aligned}
\end{equation*}

The second set of constraints, C2, avoids solutions that map on the same \a{pe} more than a single nodes at the same time, and is encoded through:
\begin{equation}
	\begin{aligned}
	    M(n,m) &= \bigwedge_{(v_i,w_j)\in \mathcal{V}(n,m)} \neg (v_i \wedge w_j) \\
		\xi &= \bigwedge_{n}^{N - 1} \bigwedge_{m = n + 1}^{N} M(n,m)\\
	\end{aligned}
	\label{eq:diffpe}
\end{equation}
with $\mathcal{V}(n,m)$ defined as:
\begin{equation*}
    \mathcal{V}(n,m) = \{(v_i,w_j): v_i \prec w_j, v_i\in \mathcal{L}(n), w_j \in \mathcal{L}(m)\}
\end{equation*}

The last set of constraints, C3, handles the dependencies in the \a{dfg}, and requires a lengthier explanation.
For each dependency, we consider only literals that are at most one iteration apart in the \a{kms}, on a neighbor \a{pe} and that respect one of the relations:
\begin{equation}
    \left\{
\begin{array}{@{}ll@{}}
    c_d \leq c_s \text{ if } it_s \neq it_d\\
    c_d > c_s \text{ if } it_s = it_d
\end{array}\right.
    \label{eq:cc}
\end{equation}
where $c_d$ is the \a{cgra}-cycle at which the destination node is scheduled, and $c_s$ is the \a{cgra}-cycle at which the source node is scheduled. 
For example, note that for $n_s = n_{10}$ and $n_d = n_{11}$, at $c_s = 1 $ and $c_d = 2$ at $it_s = 0$ and $it_d = 0$ the second relation of \autoref{eq:cc} is satisfied.
This constraint ensures that a node consumes the value produced by the predecessor in the proper order, avoiding overlapping of the same dependencies among kernel and prologue/epilogue stages.

Before proceeding in the description of C3, we define two functions that will be used to better formalize the constraints.
The \textit{neighborhood function} expresses whether the \as{pe} of two literals are neighbors, and is defined as follows:

\begin{equation}
    f_n(v_1,v_2) = \left\{
      \begin{array}{@{}ll@{}}
        2 & \text{if neighbors and}\ p_1\neq p_2 \\
        1 & \text{if neighbors and}\ p_1=p_2 \\
        0 & \text{otherwise}
      \end{array}\right.
    \label{eq:neigh}
\end{equation}

The \textit{\a{cgra}-cycle function}  returns the \a{cgra}-cycle associated to a literal, e.g. $f_c(v_i) = f_c(x_{n,p,c,it}) = c$.

C3 enforces that each data dependency is mapped on neighbors \a{pe} on the \a{cgra} and that data produced by a node on a \a{cgra}-cycle is consumed before being overwritten by another node on a subsequent \a{cgra}-cycle.
For each edge in the \a{dfg}, we generate a set of constraints $\xi$ that will be added to the \a{sat} solver, and then we take into account all the possible combinations of the same dependency in the \a{kms}: 
\begin{equation}
    \xi = \bigvee_{(n_s,n_d)\in \a{kms}} \mathcal{D}(n_s,n_d)
    \label{eq:c3intro}
\end{equation}

Now we proceed to give all the information needed to formally define C3:

\begin{equation}
    \mathcal{D}(n_s,n_d) = \left\{
      \begin{array}{@{}ll@{}}
        \gamma & \text{if}\ distance = 1 \\
        \zeta & \text{if}\ distance \neq 1 \\
      \end{array}\right.
    \label{eq:neigh}
\end{equation}

where the $\mathcal{D}(n_s,n_d)$ refers to the constraint associated to the edge $(n_s,n_d)$ that depending on the distance value can be either $\gamma$ or $\zeta$. The distance between two nodes in the \a{kms} is computed with the following equation:
\begin{equation}
    (c_d - c_s + \a{ii} ) \mod \a{ii}
    \label{eq:distance}
\end{equation}
\begin{figure}
    \centering
    \includegraphics[width= 0.4\textwidth]{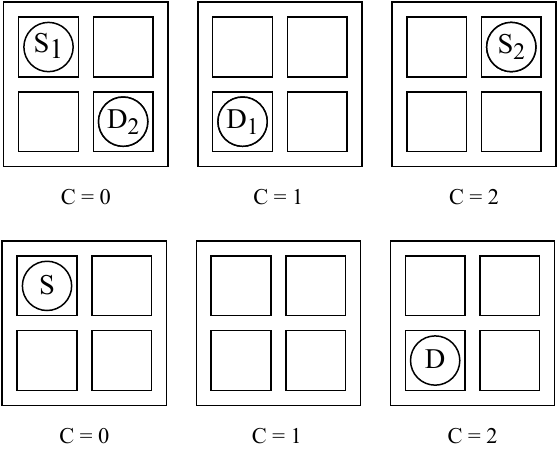}
    \caption{Top: two examples of dependencies -- $S_1$, $D_1$ and $S_2$,$D_2$ -- with a distance equal to 1.\\ Bottom:  an example of dependence with a distance greater than 1.}
    \label{fig:c3case}
\end{figure}

For example, let us consider $n_s = n_2$ of iteration 0 at \a{cgra}-cycle 0 ($c_s = 0$ and $n_d = n_9$ of iteration 0 at \a{cgra}-cycle 2 ($c_d = 2$). In this case, we have $(2 - 0 + 3 ) \mod 3 = 2$.

\autoref{eq:distance} returns the number of \a{cgra}-cycles that the source node needs to wait for, until its value is consumed. We use this information to distinguish between two main cases depicted in \autoref{fig:c3case}. The figure on the top shows the case when source and destination node, on different \as{pe}, are only one \a{cgra}-cycle apart; on the bottom, source and destination are more than one \a{cgra}-cycle apart.

We now proceed to define $\gamma$ and $\zeta$. 

\begin{equation}
    \gamma = \bigvee_{(v_i,w_j)\in \mathcal{A}(n_s,n_d)} (v_i \wedge w_j)
    \label{eq:cc1}
\end{equation}
where $\mathcal{A}(n_s, n_d)$ is a set of literals defined as:
\begin{equation}
    \mathcal{A}(n_s, n_d) = \{(v_i,w_j) : f_n(v_i,w_j) > 0, v_i\in \mathcal{L}(n_s), w_j \in \mathcal{L}(n_d)\}
    \label{eq:a}
\end{equation}
Now we define $\zeta$ as the \textit{or} between two sets of literals.
\begin{equation}
    \zeta = \zeta_1 \vee \zeta_2
\end{equation}
where $\zeta_1$ 
is defined as:
\begin{equation}
    \zeta_1 = \bigvee_{(v_i,w_j)\in \mathcal{B}(n_s,n_d)} (v_i \wedge w_j)
    \label{eq:cc1}
\end{equation}
and where $\mathcal{B}(n_s, n_d)$ is a set of literals defined as:
\begin{equation}
    \mathcal{B}(n_s, n_d) = \{(v_i,w_j) : f_n(v_i,w_j) = 1, v_i\in \mathcal{L}(n_s), w_j \in \mathcal{L}(n_d)\}
\end{equation}
Lastly we define $\zeta_2$ as:
\begin{equation}
    \zeta_2 = \bigvee_{(v_i,w_j)\in \mathcal{A}(n_s,n_d)} \left(v_i \wedge w_j \wedge \neg \left( \bigvee_{z_k\in \mathcal{V}(ns,n_d)} z_k \right)\right)
    \label{eq:cc2}
\end{equation}
where $\mathcal{A}(n_s, n_d)$ is constructed as in \autoref{eq:a} and $\mathcal{V}(n_s, n_d)$ is the set of literals with the same \a{pe} of the source node on different \a{cgra}-cycle defined as:
\begin{equation}
    \mathcal{V}(n_s, n_d) = \{z_k : f_c(v_i) < f_c(z_k) < f_c(w_j), z_k\in \mathcal{L}\}
\end{equation}
The \a{cnf} $\zeta$ is composed of two terms  because $\zeta_1$ handles the case in which the output of the source node is delivered to the destination node through the registers on the \a{pe} ($n_s$ and $n_d$ necessary on the same \a{pe}), while $\zeta_2$ handles the case in which the output is written in the output register of the \a{pe}, and hence it must not be overwritten in subsequent \a{cgra}-cycles.
\begin{figure}
    \centering
    \includegraphics[width= 0.4\textwidth]{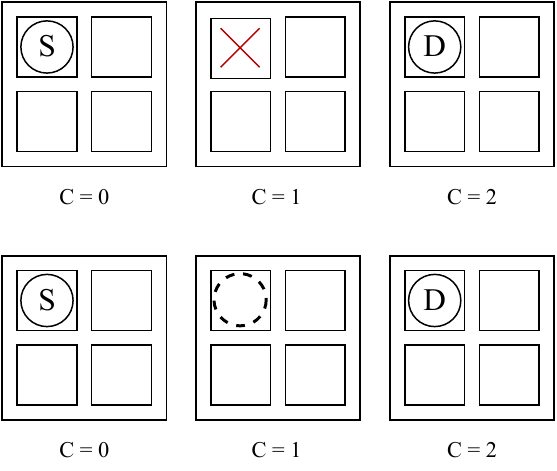}
    \caption{Top: the \a{pe} used by the source node must be left empty until the destination node consumes its value.\\ Bottom: the \a{pe} used by the source node can be also used in subsequent \a{cgra}-cycles, as the value written by the source node is, in this case, saved in a register (this will require Register Allocation to be validated).}
    \label{fig:c3case2}
\end{figure}
\autoref{fig:c3case2} shows those two cases. The figure on top shows that the \a{pe} 0 at \a{cgra}-cycle 1 cannot be filled with another instruction, since it will overwrite the data produced by $n_s$. The figure on the bottom shows instead that the \a{pe} 0 at \a{cgra}-cycle 1 can be overwritten since the destination node is on the same \a{pe} as the source node and an internal register can be used.
The \a{sat} solver will then decide which of the two possibilities is used to generate a mapping.
Giving both these options to the solver provided the key to find better mapping in several cases, as mentioned in the Experimental Section.
Lastly, back dependencies need to be handled in a slightly different way. In particular, \autoref{eq:cc} becomes:

\begin{equation}
    \left\{
\begin{array}{@{}ll@{}}
    c_d \leq c_s \text{ if } it_s = it_d\\
    c_d > c_s \text{ if } it_s > it_d
\end{array}\right.
    \label{eq:cc2}
\end{equation}

This concludes the description of the third set of constraints. 
For the running example, the satisfying solution given by the \a{sat} solver sets the following literals to true: $v_{11,1,0,0}$, $v_{6,2,0,0}$, $v_{7,3,0,0}$, $v_{2,0,1,0}$, $v_{1,1,1,1}$, $v_{8,2,1,0}$, $v_{3,3,1,1}$,  $v_{9,0,2,0}$, $v_{10,1,2,1}$, $v_{4,2,2,1}$, $v_{5,3,2,1}$.
This, in turn, corresponds to the mapping shown in \autoref{fig:mapres}.b.

Register Allocation is the last phase needed in order to validate the mapping returned by the solver.

\subsection{Register Allocation}
\label{sec:register_allocation}
In our CFN formulation, to have a more straightforward set of constraints, we chose not to include information about which register should store the data produced by the node on every PE. We hence treat the \a{ra} task as a separate problem.
Opting for a SAT formulation that yields register-oblivious solutions was a strategic decision aimed at minimizing compilation times. However, a consequence of this approach is the necessity to validate the solutions returned by the scheduling phase, in terms of register allocation needs. This issue is observable in only 3 of our experiments (\texttt{backprop} $3\times 3$, $4\times 4$, and $5\times 5$). 
Generally, after the solver returns a mapping, the RA phase needs to ensure that the CGRA has an adequate number of registers to execute the CIL correctly. SAT-MapIt, as a separate phase subsequent to SAT solving, exploits the SSA format of the input code and looks for an optimal solution\cite{hack2006optimal} of the RA problem. 
For each \a{pe} in the \a{cgra}, we extract all kind of dependencies referenced by that PE and we construct an interference graph. We then attempt to color using $n$ colors, where $n$ is the number of registers available on each PE. 

If the coloring succeeds, we know that no further action is needed, and the toolchain provides the assembly code for the \a{cgra}.
However, if the coloring process fails, we increment the II and initiate a new search. While it's possible to explore additional graph splitting techniques to make the graph colorable, we do not consider them herein, as such optimizations are orthogonal to the mapping problem we address in this work.

\section{Experimental Results:  \toolname{} solutions}
\label{sec:results}
In this section, we evaluate the effectiveness of \toolname\ on a set of \as{cil} from MiBench and Rodinia benchmark suites, shown in \autoref{tab:benchlst}. Table \autoref{tab:benchlst} provides more insights into the types of applications that are accelerated in our experiments. It also highlights how our methodology is applied to single loops, often focusing on the innermost ones, which were extracted utilizing the RAMP toolchain. The table includes information on the number of nodes, edges, and lines of code for each benchmark. It is important to note that these metrics alone do not fully capture the complexity involved in mapping a given loop, but they still enrich the understanding of the applications' complexity.

We compare the obtained $\a{ii}$, and the time to find it,  with respect to two techniques of the \a{soa}: RAMP\cite{dave2018ramp} and PathSeeker \cite{pathseeker}. We show results for only those benchmarks where RAMP and PathSeeker were able to provide a solution and where CFG was not present within the CIL body.  
\begin{table}
\caption{List of benchmarks with graph information}
\label{tab:benchlst}
\begin{tabular}{c|l|ccc}
\hline
\multicolumn{1}{l|}{\textbf{Suite}} & \textbf{CIL} & \textbf{\#nodes} & \textbf{\#edges} & \textbf{\#lines of code}\\ \hline
\multirow{7}{*}{\texttt{MiBench}}     & \texttt{sha}             & 30             & 33      & 1            \\ 
                                      & \texttt{sha2}            & 26             & 28      & 1            \\ 
                                      & \texttt{gsm}             & 20             & 24      & 3            \\ 
                                      & \texttt{patricia}        & 42             & 46      & 3            \\ 
                                      & \texttt{bitcount}        & 26             & 29      & 1            \\ 
                                      & \texttt{basicmath}       & 19             & 20      & 6            \\ 
                                      & \texttt{stringsearch}    & 16             & 16      & 2            \\ \hline
\multirow{4}{*}{\texttt{Rodinia}}     & \texttt{backpropagation} & 35             & 39      & 1            \\ 
                                      & \texttt{nw}              & 16             & 16      & 1            \\ 
                                      & \texttt{srand}           & 22             & 22      & 3            \\ 
                                      & \texttt{hotspot}         & 67             & 76      & 8            \\ \hline
\end{tabular}
\end{table}

\subsection{Experimental Setup}
\label{sec:setup}
We use the original code publicly released by the authors of the two \a{soa} tools and the same \a{dfg}, while in performance evaluation we extract the \a{dfg} directly from the C code and apply an instruction selection pass to convert the LLVM-IR instructions into instructions compatible with the \a{isa} of the \a{cgra} used.
In the target \a{cgra} architecture that we consider in our experiments, each \a{pe} is connected to the four nearest neighbors, as in \autoref{fig:cgra_arch}, and each \a{pe} contains four local registers. We vary the size of the mesh from $2\times2$ up to $5\times5$.
The Z3 solver \cite{moura2008z3} is used to solve our \a{sat} formulation. All experiments are performed on a machine with 2.6 GHz 6-Core Intel Core i7. For PathSeeker, each experiment was repeated 10 times.

\subsection{\toolname\ achieves better \a{ii}}\label{sec:betterII} 
\begin{figure*}[htp]
\centering
\includegraphics[width=\textwidth]{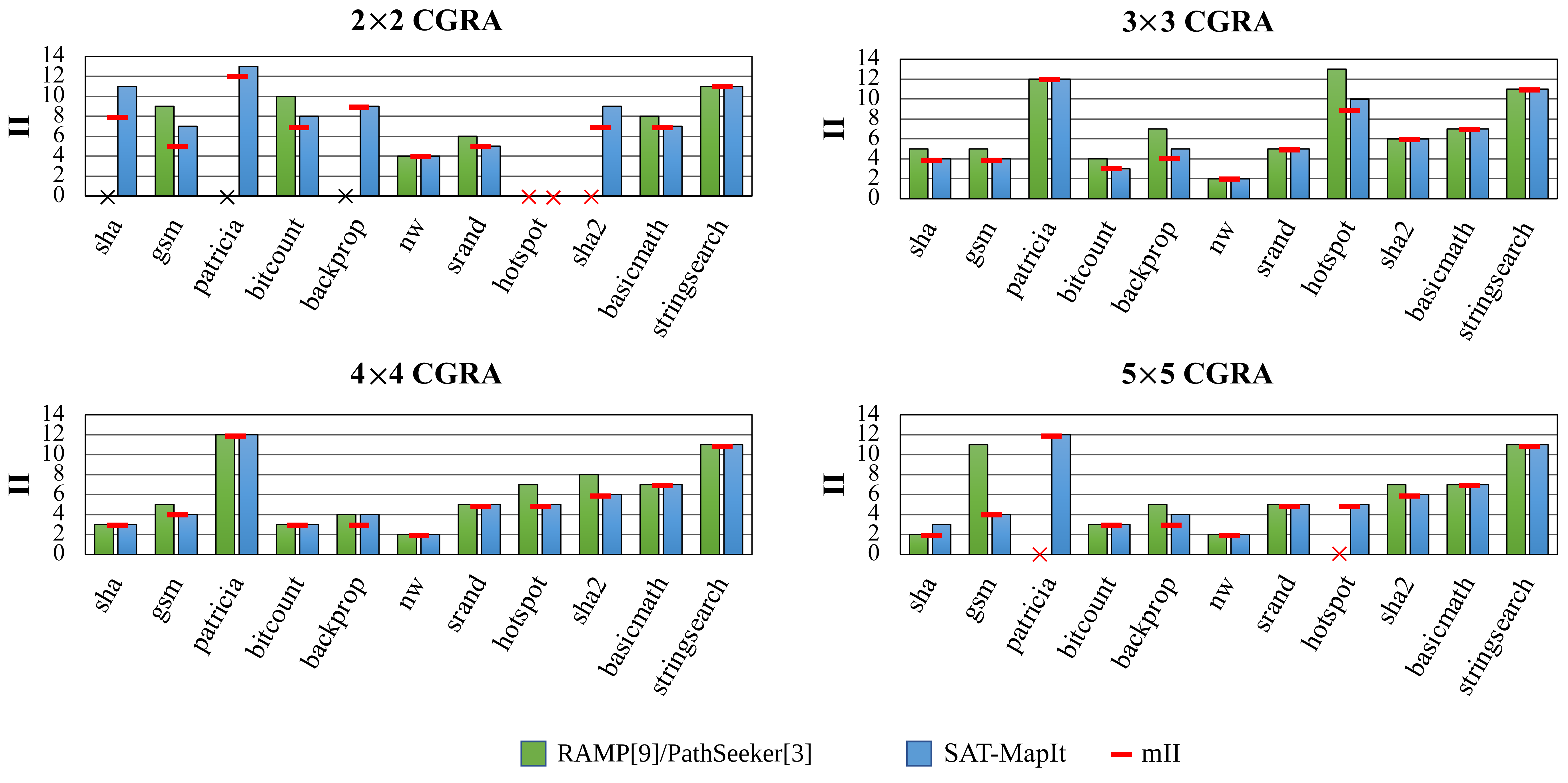}\hfill

\caption{Experimental results of the chosen benchmarks for different architecture sizes.
We compare the \a{ii} found by \toolname\ with respect to the best results obtained by RAMP and PathSeeker -- lower is better. A red cross means that the process did not terminate before a timeout of 4000 seconds. 
A black cross means that the process was terminated when it reached a current \a{ii} of 50 but still found no feasible solution.
Red dashes indicate the minimum II (mII) for that \glsentryshort{cil}. For the $2\times2$ \a{cgra}, \texttt{hotspot} had a \a{mii} of 17, which is not displayed.
}
\label{fig:benchs}

\end{figure*}

The performance of a mapping is first and foremost measured by the $\a{ii}$ achieved, because this, in turn, is a measure of the level of parallelism obtained; our first experiments hence compare the \a{ii} of \toolname\ with those of the \a{soa}, for each benchmark explored. This is depicted in \autoref{fig:benchs}, which shows the performance obtained by all techniques for different \a{cgra} sizes. For \a{soa}, we report the best result returned by the two algorithms.

It can be seen from \autoref{fig:benchs} that \toolname\ sets a new benchmark in the field as it almost always either outperforms the \a{soa} or achieves the II lower bound (\a{mii}), for the evaluated \as{cil}. In particular, \toolname\ found a solution in 6 out of the 7 cases where the \a{soa} could not; it reached a lower \a{ii} than the other tools for 14 \as{cil} and reached the \a{mii} (computed from \autoref{eq:mii}) for 34 \as{cil} while the \a{soa} could only achieve so for 19 of them. 
This indicates that \toolname~ explores the solution space more effectively than \a{soa} techniques, and this effect is especially visible when the mapping problem becomes more challenging, as detailed in \secref{sec:tight_resources}.
The three exceptions to these patterns are \texttt{hotspot} in the $2\times2$ \a{cgra}, where no tool could find a solution for an \a{mii} of 19 before the timeout of 4000 seconds; \texttt{backprop} in the $4\times4$ \a{cgra}, where \toolname\ matches the \a{soa} but without reaching the \a{mii}; and \texttt{sha} in the $5\times5$ \a{cgra}. We further comment on this last result in \secref{sec:limitations}.\\ 

It is worth noting that, in some benchmarks, expanding the array size fails to reduce the II (e.g. \texttt{stringsearch}). This arises from the presence of a long dependency chain within the DFG. Such chain acts as a bottleneck, limiting the potential for performance improvement through increased array size.


\subsection{\toolname\ uses tight resources better}\label{sec:tight_resources} By focusing on the $2\times2$ size \a{cgra}, it can be seen not only that \toolname~ outperforms \a{soa} in 8 of 11 benchmarks, but also that it can find a valid solution three times (sha, patricia and backpro) where \a{soa} could not. This showcases the effectiveness of our methodology, particularly when the mapping problem becomes more challenging, always managing to use the minimal number of resources possible. The other tools evaluated either return no solution, or seem to be biased towards adding routing nodes even when they are not necessary in order to find a solution. Our tool can then fully exploit and handle the resources available, even when these are tightly constrained e.g. in low-power edge architectures \cite{DuchFeb19}.

\begin{table}[bth]

\caption{Mapping time (seconds) for different CGRA sizes}
\resizebox{\textwidth}{!}{\begin{tabular}{l|ccc|ccc|ccc|ccc}
\hline
 & \multicolumn{3}{c|}{\textbf{$2\times2$}}& \multicolumn{3}{c|}{\textbf{$3\times3$}}& \multicolumn{3}{c|}{\textbf{$4\times4$}}& \multicolumn{3}{c}{\textbf{$5\times5$}} \\
\textbf{CIL} & \textbf{SoA} & \textbf{Ours}  & \textbf{$\Delta$} & \textbf{SoA} & \textbf{Ours}  & \textbf{$\Delta$}& \textbf{SoA} & \textbf{Ours}  & \textbf{$\Delta$}& \textbf{SoA} & \textbf{Ours}  & \textbf{$\Delta$} \\ \hline
\texttt{sha}            & 41.01     & 3.22   & -37.79       & 0.23  & 2.86   & 2.63       & 32.04    & 7.23     & -24.81      & 6.25     & 28.93   & 22.68      \\       
\texttt{gsm}            & 2.81      & 1.25   & -1.56        & 4.14  & 4.35   & 0.21       & 32.04    & 10.46    & -21.58      & 0.29     & 21.4    & 21.11      \\   
\texttt{patricia}       & 1351      & 5.39   & -1345        & 48.98 & 16.31  & -32.67     & 421.05   & 39.28    & -381.77     & 4000     & 75.16   & -3924   \\   
\texttt{bitcount}       & 2.63      & 1.68   & -0.95        & 5.86  & 7.84   & 1.98       & 0.44     & 21.55    & 21.11       & 0.01     & 47.62   & 47.61      \\   
\texttt{backprop}       & 1262      & 3.39   & -1259        & 44.62 & 12.27  & -32.35     & 57.17    & 25.1     & -32.07      & 981.73   & 52.43   & -929.3     \\   
\texttt{nw}             & 0.01      & 0.56   & 0.55         & 0.03  & 1.56   & 1.53       & 0.08     & 3.63     & 3.55        & 0.02     & 7.75    & 7.73       \\
\texttt{srand}          & 0.32      & 1.15   & 0.83         & 0.09  & 3.9    & 3.81       & 0.25     & 8.79     & 8.54        & 0.02     & 21.64   & 21.62      \\
\texttt{hotspot}        & 4000      & 4000   & 0            & 13.19 & 28.43  & 15.24      & 3556.62  & 3734     & 178.15      & 4000     & 108.02  & -3891   \\
\texttt{sha2}           & 4000      & 2.21   & -3997        & 61.7  & 3.13   & -58.57     & 696.6    & 7.19     & -689.41     & 675.12   & 16.88   & -658.24    \\   
\texttt{basicmath}      & 0.01      & 0.62   & 0.61         & 0.07  & 1.9    & 1.83       & 0.22     & 4.11     & 3.89        & 0.5      & 8.7     & 8.2        \\   
\texttt{stringsearch}&  0.19        & 1.02   & 0.83         & 3.27  & 3.55   & 0.28       & 0.02     & 7.62     & 7.6         & 0.02     & 15.3    & 15.28      \\   \hline
\end{tabular}}

\label{tab:times}
\end{table}

\subsection{Runtime Analysis}
Given that the $\a{ii}$ found are better than \a{soa}, it may be interesting to also analyze the time needed to find such solutions;
this is reported in \autoref{tab:times}. The following statements can be made when looking at these numbers. 1)  Our tools running time is longer than \a{soa} in 26 out of 44 experiments. 
2) However, in these 26 cases the average time difference  is only 15.28 seconds, with a standard deviation of $34.97$. 3) On the other hand, in the 18 cases in which our tool is faster, the average time difference is 962.24 seconds, with a standard deviation of $1438.78$. This shows that, in these experiments, SAT-MapIt is significantly faster than state-of-the-art methods, particularly when the problem is challenging and hence computation times are high. The solver time will, of course, still increase worse-case
exponentially with the size of the input, and it might not scale for some instances; however, where it does scale, it obtains a solution which is found on average tangibly faster than state of the art.

\subsection{Limitations of \toolname}
\label{sec:limitations}

Currently, our tool does not apply any routing strategies. This limitation manifests in the \texttt{sha} \a{cil} of a $5\times5$ \a{cgra}, where we achieve an $\a{ii}$ of 3, while \cite{dave2018ramp} and \cite{pathseeker} can find an $\a{ii}$ of 2 by adding a routing node. This is the only case, out of the 44 experiments shown, where the effect of this limitation can be noticed. 
Implementing routing capabilities in \toolname\ will be the subject of future work. 
Another limitation can be seen in the optimality of SAT-MapIt, hence in its capability of returning \textit{only} optimal solutions. Indeed, when an exact strategy does not scale (which is necessarily the case for some input instances, given  the high complexity of the problem tackled) a non-exact algorithm could provide a non-guaranteed-optimal but good-enough feasible solution.  
An non-exact strategy for SAT-MapIt could involve restricting the solver's exploration time for each II, hence potentially yielding sub-optimal yet feasible solutions. For instance, in our hotspot benchmark for a $2\times 2$ CGRA setup, no solutions emerged after a 4000 seconds (and even, in an additional experiment,  after a 24-hour timeout). However, when we imposed an exploration limit of 5 minutes and 1 minute per II 
we detected a feasible solution at $II=21$ under both conditions. These adjustments led to total compilation times of 975.3 seconds with a 5-minute limit, and 257.2 seconds with a 1-minute limit. The exploration of non-exact strategies for SAT-MapIt is left to future work.
A further limitation consists of the separation of the phases of scheduling and register allocation. The separation was a deliberate choice aimed at balancing the intricacy and computational efficiency of the problem-solving process. Our decision to decouple register allocation from the SAT formulation and address it in a subsequent phase is rooted in the desire to maintain clarity and manageability within the problem formulation. However, we acknowledge that this decoupling might lead to suboptimality, and this represents a potential direction for future work.

\subsection{An example of mapping}

\begin{figure}[htp]
\centering
\includegraphics[width=0.48\textwidth]{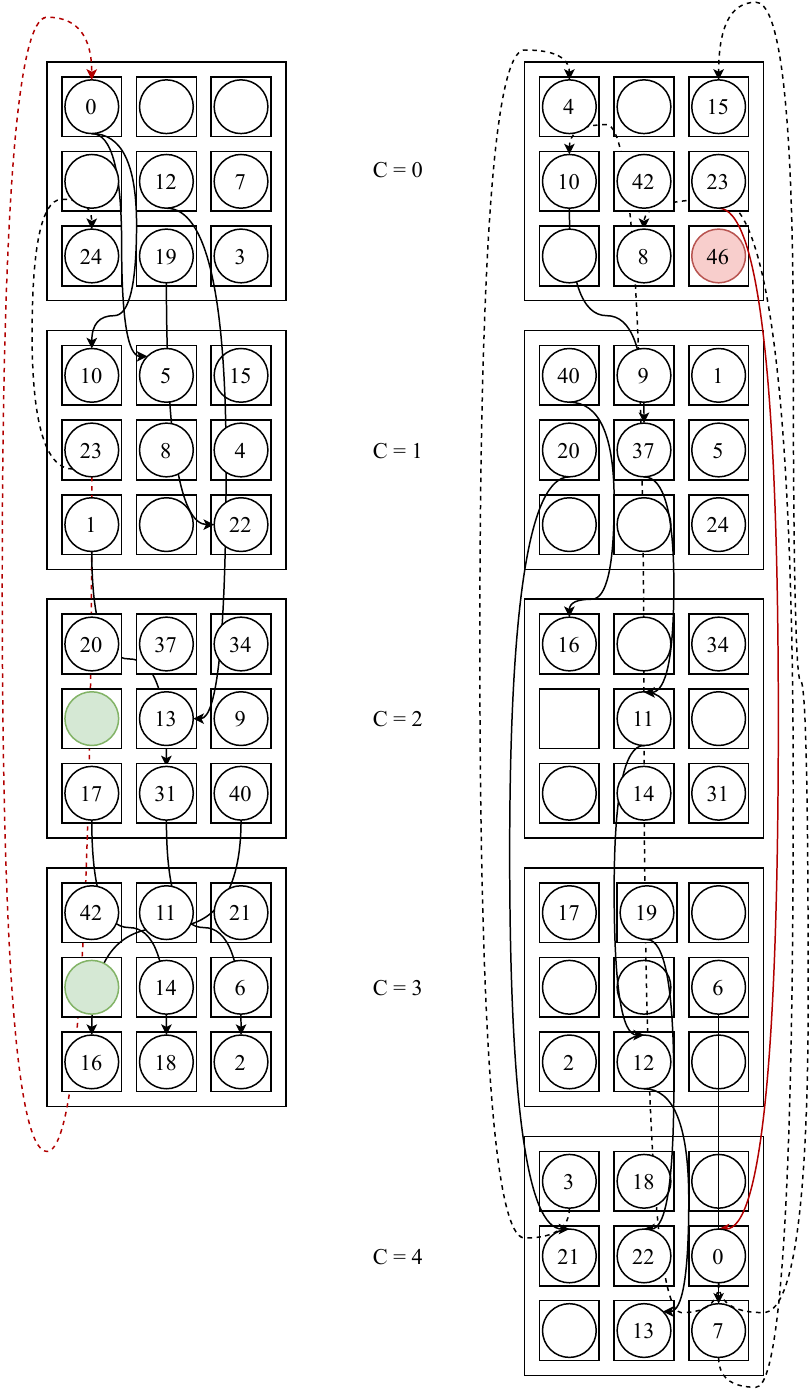}\hfill
\caption{Mapping results of \texttt{sha1} \a{cil} for both tools. RAMP on the right, \toolname~ on the left}
\label{fig:mapex}

\end{figure}

Lastly, we show in more detail one of the mappings where we achieved a better \a{ii} than \a{soa}:  benchmark \texttt{sha1} and \a{cgra} size $3\times3$. \autoref{fig:mapex} shows the two mappings obtained for this benchmark: to the left, by \toolname, and to the right by RAMP. The \a{dfg} contains 30 nodes and 33 edges -- all nodes and only a subset of the edges are represented in the picture.
In the description of our methodology, and in particular in how to handle dependencies, we showed in \autoref{fig:c3case2} how \toolname\ can handle two different cases of mappings. This capability will be showcased here, considering in particular a dependence from node 23 to node 0, and then another one from node 12 to node 13.
 We can see that the PE3 is left empty until the data is consumed, which is one of the options available to the \a{sat} solver and encoded through \autoref{eq:cc2}. Instead, consider now the dependence between node 12 and node 13. In that case, we can see that the constraint associated with it is the one in \autoref{eq:cc1}, which allows node 8 to be mapped on the same \a{pe} of node 12, exploiting the internal registers of \a{pe} 4.
With these capabilities, \toolname\ finds a lower $\a{ii}$, specifically $\a{ii} = 4$. While RAMP returns a mapping with $\a{ii} = 5$. We can also notice that node 46 (in red) is an unnecessary routing node that RAMP added to resolve a dependency and map the \a{cil}.

\section{Application-to-hardware framework}
\label{sec:arch}




\begin{figure*}
\centering
\minipage[t][1cm][t]{0.85\textwidth}
  \includegraphics[width=\linewidth]{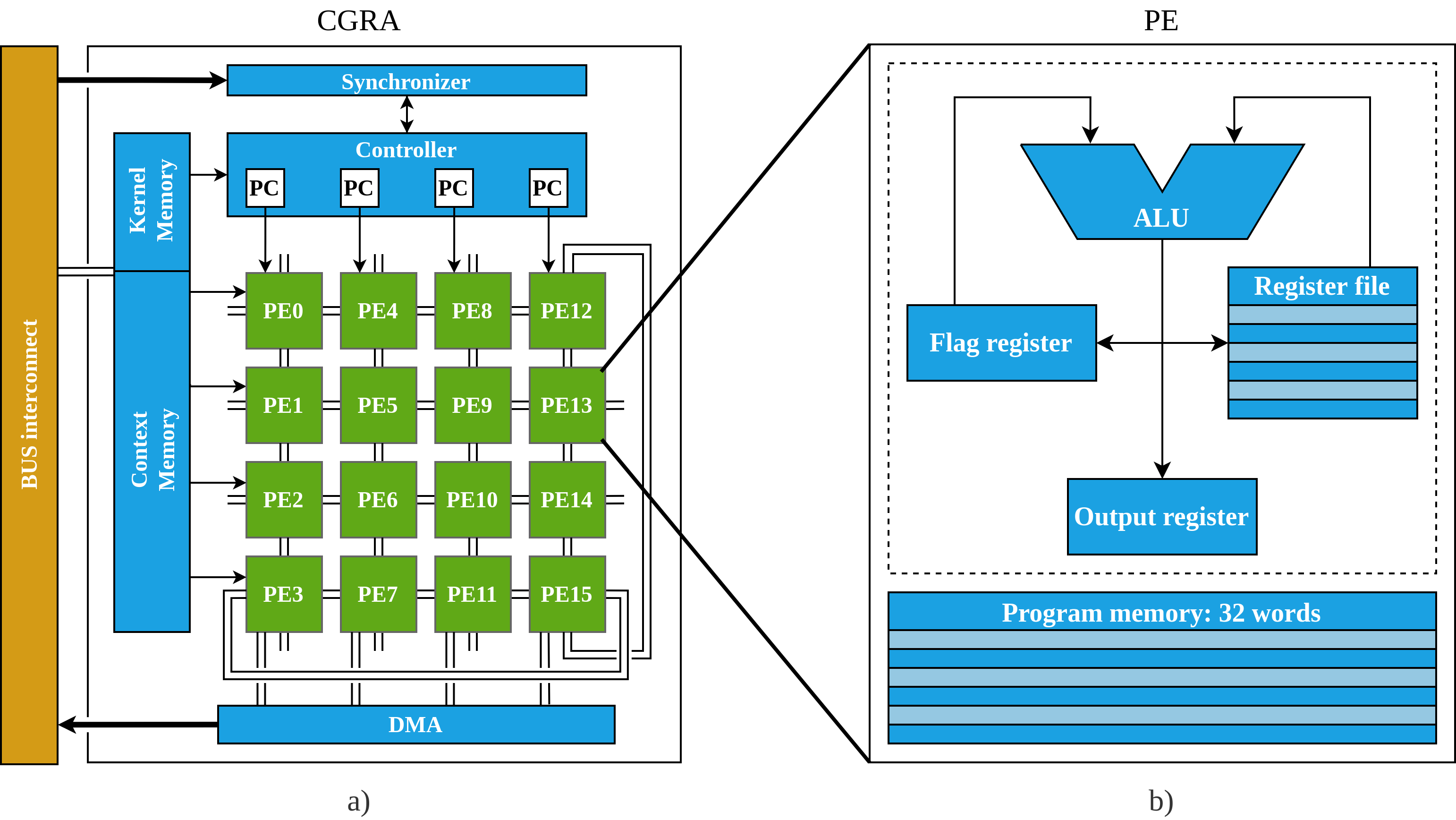}
\endminipage\hfill
\caption{a): Top-level view of the \textit{OpenEdgeCGRA} architecture with a $4\times4$ \a{pe} array, b): \a{pe}-level architectural view}
\label{fig:cgra_arch}
\end{figure*}

\toolname{} operates on an abstract view of applications and architectures. In particular, it relies on an instructions-level representation  of \as{cil}, according to the single-assignment form provided by the LLVM \a{ir}. Such hardware-agnostic stance allows to flexibly target  any \a{cgra} architecture supporting modulo scheduling \a{isa}~\cite{DuchFeb19, singh2000morphosys}. 
Nonetheless, hardware-specific back-ends must be employed to translate the compiler output to a binary representation compatible with a target architecture. After back-end translation, the resulting bitstream contains a sequence of control words, dictating the desired behavior of the \as{pe} at \a{cil} execution time, for each \a{cgra}-cycle. These instructions encode the operation performed by \as{alu}, the input operands, and the output destination. 
Note that, in modulo scheduled \as{cgra}, several instructions are configured in each \a{pe}, with one being executed in all \as{pe} at every \a{cgra}-cycle, to implement modulo scheduling as discussed in \secref{sec:background}.
Individual bits of instructions directly or indirectly drive the multiplexers that are responsible for the control logic.

A direct connection between the control words and multiplexers is featured in \textit{OpenEdgeCGRA} \cite{OpenEdgeCGRA}, the \a{cgra} considered in this work, waving the need for a decoder for faster \a{cil} execution. In particular, we target the implementation available in the public repository in ~\cite{RodriguezMay2023}. It is fully synthesized and can be simulated at the post-synthesis level, allowing us to inspect the performance and energy requirement of mapped \as{cil}, which are presented in \secref{sec:results}.

The \textit{OpenEdgeCGRA} architecture, shown in \autoref{fig:cgra_arch}, comprises a two-dimensional array of \as{pe}, interconnected with their nearest neighbors in a torus,  wrapping around columns and rows (\autoref{fig:cgra_arch}.a). \as{pe} are organized into columns, each having its own Program Counter (PC). 
At the array periphery, a \a{dma} engine interfaces \as{pe} with the system bus, providing one input/output port per \a{cgra} column. A context memory acts as a cache for configurations, storing the bitstreams encoding the required \as{cil}. The descriptor of the \a{cil} (index, required number of columns and location of the configuration words in the context memory) are stored in the Kernel Memory. Each \a{pe} is composed of data registers, a program memory and an \a{alu} (\autoref{fig:cgra_arch}.b). The program memory governs the local behavior of the \a{pe} according to the value of the word indexed by the PC. Each program memory word is composed of 32 bits, and encodes the source of operands (immediate, register file, neighboring cells or main memory via the system bus), where the output is stored, and the operation being performed. 


\as{alu} can perform arithmetic operations (signed addition, subtraction and multiplication, as well as fixed-point multiplication), arithmetic and logic shifts, and bit-wise operations. The \as{alu} also supports selecting operands based on the state of the zero and sign flags (BXFA and BSFA, respectively), thus allowing the implementation of branches via if-conversion. Data transfers to/from memory are realized by load and store opcodes. Finally, \as{pe} can conditionally or unconditionally overwrite the value of program counters by issuing branch and jump instructions. 
Such instructions are used to govern mapped modulo-scheduled \as{cil}, so that the prologue is executed once, the kernel is iterated as required by the loop count, before ending with the epilogue. Local \a{pe} data storage is provided by a 4-word register file, the \a{pe} output register, and a flag register storing the 1-bit zero and sign flags resulting from the previous instruction. 

The \a{cgra} \a{isa} closely mimics the integer set of opcodes supported by LLVM, as reported Table \ref{tbl:instr_set}. This characteristic results in a simple back-end implementation. On the other side of the coin, it restricts its support to integer \as{cil}. Hence, we only considered integer benchmarks for the exploration results discussed in \secref{sec:results_hardware}.

\begin{table}
\caption{Instructions set architecture of the open hardware \a{cgra} \a{isa}, as reported in \cite{RodriguezMay2023}.}
\label{tbl:instr_set}
\centering
\begin{tabular}{@{\hskip0.1cm}l@{\hskip0.1cm}@{\hskip0.1cm}l@{\hskip0.1cm}} \hline
\textbf{Type of instruction} & \textbf{Opcode} \\ \hline
Arithmetic operations & SADD, SSUB, SMUL, FXPMUL \\
Shifts  & SLT, SRT, SRA \\
Bit-wise operations  & LAND, LOR, LXOR, LNAND, LNOR, LXNOR \\
Selects  & BSFA, BZFA \\
Loads and stores  & LWD, LWI, SWD, SWI \\
Conditional and unconditional branches & BEQ, BNE, BLT, BGE, JUMP \\
No operation & NOP \\
Finish  & EXIT \\
\hline
\end{tabular}
\end{table}


At run-time, when a \a{cil} execution is requested, its descriptor is fetched from the Kernel Memory, and the synchronizer module checks if enough \a{pe} columns are idle. If this is not the case, it waits for enough resources to become available, otherwise it copies the configuration words from the context memory to the program memory of \as{pe}. \as{cil} are then executed according to the selected configurations, governed by the value of program counters. Execution ends when a \a{pe} issues the exit instruction.



\section{Experimental results: run-time performance of compute-intensive loops mapped with \toolname{}}\label{sec:results_hardware}

In this section, we investigate the characteristics of mapped \as{cil} from a hardware perspective, showcasing the effect of provisioning different amounts of resources on performance and energy efficiency. Bridging compiler-level and hardware-level explorations, we then inspect the degree of correlation between respective metrics in the two spaces: \glsentrylong{ii} and \glsentrylong{u} in the former case;  Energy, Latency, in the latter one.

\subsection{Experimental Setup}\label{sec:pi_setup}

We consider square \textit{OpenEdgeCGRA} instances with sizes $2\times2$, $3\times3$ and $4\times4$, referred to as $D2$, $D3$ and $D4$, respectively, in the following.
Benchmark \as{cil} from the MiBench suite were executed on the architectures. To compare with the \a{soa} in \secref{sec:betterII} we used the same \as{dfg} generated by the RAMP toolchain; however those \as{dfg} are not inherently compatible with the target \a{cgra}'s \a{isa}, preventing the direct execution of experiments using identical graphs. To address this discrepancy, we recalculated the \as{dfg} for each \a{cil} to align with the instruction set of \textit{OpenEdgeCGRA}, for this part of the experiments.Table \ref{tab:new_dfg} shows the number of nodes and edges in the newly generated DFGs. Consequently, this resulted in different \a{ii} values compared to \autoref{fig:benchs}.
\begin{table}[t!]
\centering
\caption{List of benchmarks with graph information after the Instruction Selection phase targeting OpenEdgeCGRA}
\begin{tabular}{l|l|l}
\hline
\multicolumn{1}{c|}{\textbf{Benchmarks}} & \multicolumn{1}{c|}{\textbf{\#nodes}} & \multicolumn{1}{c}{\textbf{\#edges}} \\ \hline \hline
reversebits                              & 9                                     & 10                                    \\ \hline
bitcount                                 & 6                                     & 7                                     \\ \hline
sqrt                                     & 8                                     & 12                                    \\ \hline
stringsearch                             & 16                                    & 18                                    \\ \hline
gsm                                      & 14                                    & 20                                    \\ \hline
sha                                      & 25                                    & 29                                    \\ \hline
sha2                                     & 25                                    & 33                                    \\ \hline
\end{tabular}
\label{tab:new_dfg}
\end{table}

Because execution time can vary across runs based on input data, all inputs were provided from a 32-bit pseudo-random number generator adapted from \cite{lecuyer1999}. Furthermore, where the array lengths could vary in each run, these were fixed as reported in \autoref{tab:energy_latency}. Across sizes, \as{cil} received the same series of inputs to allow a fair comparison. 

Experimental results were collected through postsynthesis simulations for the \technology{} process. In such technology, the $D4$ \a{cgra} required an area of $\sim\SI{0.4}{\milli\meter\squared}$. In all cases we considered a clock frequency of \SI{100}{\mega\hertz} and a voltage supply of \SI{1.2}{\volt}. \as{cil} were executed 100 times each and the results were averaged.

\begin{table}
\centering
\caption{Compiler level (\a{ii}, \a{u}) and run-time level (Energy, Latency) performance of benchmark \as{cil}.}
\begin{threeparttable}

\begin{tabular}{l||ccc|ccc||ccc|ccc}
\hline
\textbf{ }&\multicolumn{3}{c|}{\textbf{\a{ii}}}&\multicolumn{3}{c||}{\textbf{\a{u} (\%)}}&\multicolumn{3}{c|}{\textbf{Energy (nJ)}}&\multicolumn{3}{c}{\textbf{Latency (clock cycles)}}\\
\textbf{\a{cil}}&\textbf{$D2$}&\textbf{$D3$}&\textbf{$D4$}&\textbf{$D2$}&\textbf{$D3$}&\textbf{$D4$}&\textbf{$D2$}&\textbf{$D3$}&\textbf{$D4$}&\textbf{$D2$}&\textbf{$D3$}&\textbf{$D4$}\\
\hline
\texttt{reversebits}&3&3&3&75&33&19&1.5&2.1&2.8&155&158&159\\
\texttt{bitcount}&4&4&3&38&17&15&1.0&1.3&1.6&121&122&110\\
\texttt{sqrt \tnote{⁎}}&5&5&5&40&18&10&1.5&1.9&2.7&174&174&176\\
\texttt{stringsearch \tnote{\textdagger}}&4&2&2&100&89&50&13.5&5.1&6.5&1,150&349&351\\
\texttt{gsm \tnote{\textdaggerdbl}}&5&3&4&70&52&22&4.2&4.7&7.0&389&326&388\\
\texttt{sha}&7&4&3&89&69&54&13.5&13.4&14.4&1,150&862&653\\
\texttt{sha2}&–&8&8&–&36&20&–&4.7&6.0&–&289&289\\
\bottomrule
\end{tabular}
\begin{tablenotes}
\item[⁎] \footnotesize Input values are limited to unsigned 31 bits.
\item[\textdagger] \footnotesize Input values are limited to [1, 255]. Pattern length \textit{patlen} is set to 50 words. Variable \textit{skip2} is fixed to an arbitrary constant value.  
\item[\textdaggerdbl] \footnotesize Input values are limited to signed 16 bits. Array length is set to 40 words.

\end{tablenotes}
\end{threeparttable}
\label{tab:energy_latency}
\end{table}

\subsection{Experimental Results}\label{sec:pi_results}

    For every \as{cil}, we compare the execution time and energy of different CGRA sizes against the execution of the ultra-low power RISC-V microcontroller from the X-HEEP framework presented in~\cite{machetti2024x}. X-HEEP is an open-source, configurable and extendable platform. In the considered implementation, which was taped-out in \cite{machetti2024x}, it integrates the lightweight \texttt{cv32e2} core, running at  \SI{100}{MHz}, with a power consumption during processing of \SI{900}{\micro\watt}, as derived from~\cite{Carpentieri2024}. For the different \as{cil} and sizes, the powered consumed by the \a{cgra} varies between $1\sim{}2\times$ that of the X-HEEP's CPU. This means that even a small speed-up can translate into a decrease on overall energy. \autoref{tab:speedup} shows how energy gains can be obtained from executing most \as{cil} on the OpenEdgeCGRA when compared to the CPU.  
    
    Control intensive loops, or those with short kernels, pose a high strain on the CPU which needs to frequently perform counter increments, comparisons and jumps. On a CGRA, these tasks can be parallelized leading to speed-up of up to $3.8 \times$ among the benchmarked CILs. Other works have shown how CILs which make an intensive use of the memory can also greatly benefit from execution on the OpenEdgeCGRA~\cite{Carpentieri2024}.

    The \texttt{sha} kernel is the only case in which using the \a{cgra} always translates into energy increase. Its access to discontiguous positions in memory requires the \a{cgra} to recompute addresses on every iteration, demanding a large number of multiplications for which the ISA is not optimized. 

    One additional note-worthy observation from Table \ref{tab:speedup} is that the $D4$ CGRA is never the most energy-efficient, and that in some cases a trade-off between speed-up and energy exists (e.g. \texttt{bitcount, sha}). This underlines both the suitability of smaller CGRAs to accelerate edge kernels and the relevance of a translation between compiler-level and runtime metrics.

        \begin{table}[t!]
\centering
\caption{Gains obtained from executing the benchmark CILs on the CGRA vs. executing them on the open-hardware X-HEEP platform from \cite{machetti2024x}. }
\begin{threeparttable}

\begin{tabular}{l|ccc|ccc}
\toprule
\textbf{ }&\multicolumn{3}{c|}{\textbf{Speed-up}}&\multicolumn{3}{c}{\textbf{Energy gain}}\\
\textbf{Kernel}&\textbf{$D2$}&\textbf{$D3$}&\textbf{$D4$}&\textbf{$D2$}&\textbf{$D3$}&\textbf{$D4$}\\
\hline
\textbf{reversebits}&3.8×&3.7×&3.6×&3.5×&2.5×&1.8×\\
\textbf{bitcount}&1.7×&1.6×&2.0×&1.8×&1.3×&1.3×\\
\textbf{sqrt}&1.7×&1.7×&1.6×&1.8×&1.4×&1.0×\\
\textbf{stringsearch}&2.3×&3.7×&3.7×&1.8×&2.3×&1.8×\\
\textbf{gsm}&1.4×&1.9×&1.7×&1.2×&1.2×&0.8×\\
\textbf{sha}&1.0×&1.4×&1.8×&0.7×&0.8×&0.7×\\
\textbf{sha2}&N/A&2.4×&2.4×&N/A&1.3×&1.0×\\
\bottomrule
\textbf{Average}&2.0×&2.4×&2.4×&1.8×&1.6×&1.2×\\
\bottomrule
\end{tabular}

\end{threeparttable}
\label{tab:speedup}
\end{table}
    
Obtained run-time latency and energy values across benchmarks and architectures are summarized in  \autoref{tab:energy_latency}. In addition, the table also reports compiler-level metrics: \a{ii} and U. Experimental results are discussed in the following, providing insights on the inter-relationship among architectural parameters (\a{cgra} size), compiler-level metrics and run-time ones.

\paragraph{Relationship between energy and size.}
It can be observed that smaller \as{cgra} are more energy efficient in those cases in which increasing the \a{cgra} size does not result in a decrease in \a{ii}, as seen in \autoref{tab:energy_latency} for \as{cil} \kreve{} and \ksqrt{} for all sizes, where the $D2$ \a{cgra} has the smallest energy envelope. 
This amplifies the importance of a compiler that is able to perform scheduling not just when real estate in the \a{cgra}s is abundant, but even when the resource budget is tight, such as the compiler presented here: the most important energy savings are achieved in this case.

\begin{figure}[t]
\centering
\includegraphics[width=0.9\linewidth]{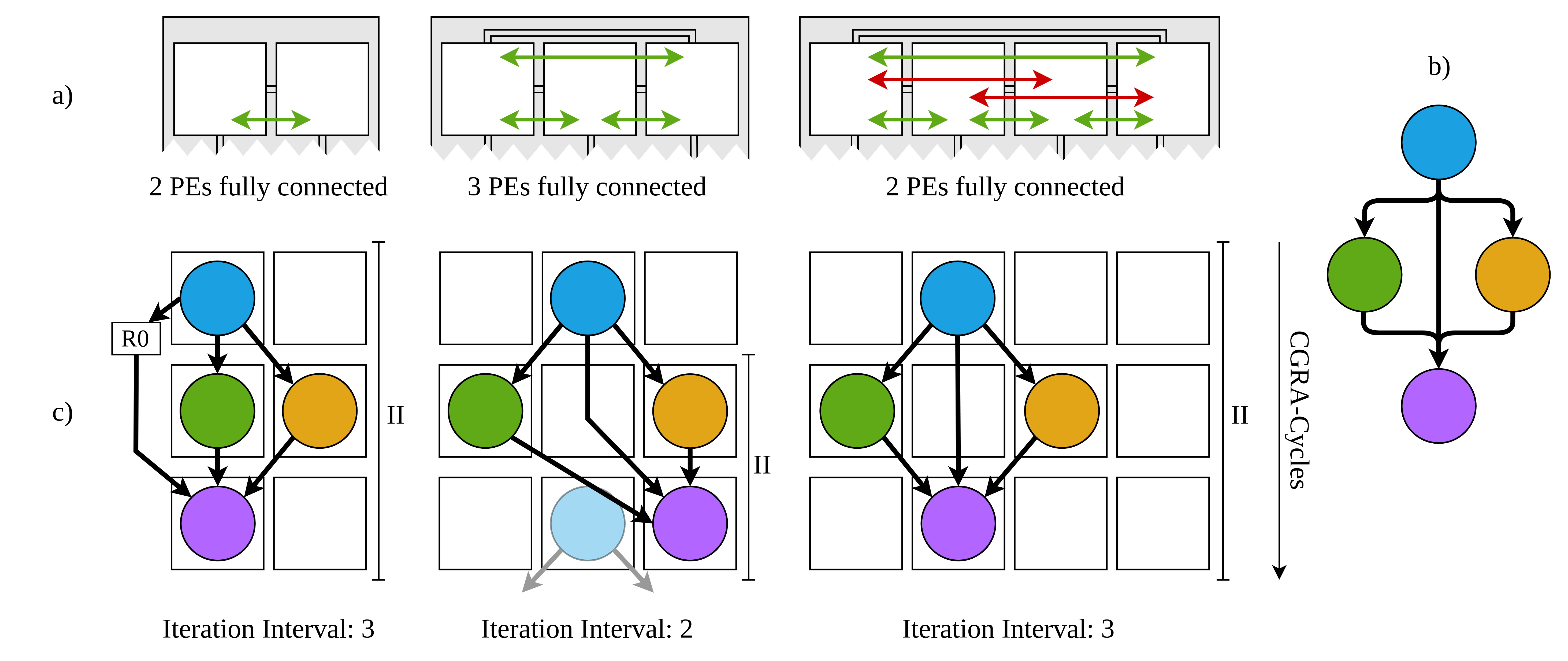}
\caption{a: In a 2-\as{pe} and 3-\as{pe} row (or column), processing elements are fully connected in a torus mesh. Adding an extra \a{pe} reduces the number of fully connected \as{pe}. b: An example \a{dfg} where full connectivity leads to a lower II. c: The mapping of the \a{dfg} from b) across different \a{cgra} dimensions. Each row represents a \a{cgra}-cycle on one row of \as{pe}. An II of 2 cycles is only achievable if 3 fully connected \as{pe} are available.}
\label{fig:wraparound}
\end{figure}

\paragraph{Relationship between \a{ii} and size}
In almost all \as{cil}, \a{ii} has a monotonic relationship with \a{cgra} size. The \kgsm{} \a{cil} is an exception where $D3$ has the best \a{ii}. In \textit{OpenEdgeCGRA}, this counter-intuitive scenario is found when the mapping exploits the wrap-around between the \as{pe} of the edge. When the \a{cgra} size increases, this connection is lost and a less efficient mapping is achieved. 
If nodes in a section of the \a{dfg} are fully connected (all nodes are related to each other), they can be efficiently mapped in a set of \as{pe} that are also fully connected. \autoref{fig:wraparound}.a illustrates how the $D3$ \a{cgra} offers the highest number of fully connected \as{pe}. Therefore, if the sample \a{dfg} from \autoref{fig:wraparound}.b is mapped into the three \a{cgra} dimensions, the best \a{ii} that can be obtained with \toolname{} is in $D3$. Such mapping is shown in \autoref{fig:wraparound}.c along with its counterparts for $D2$ and $D4$. 

\paragraph{Relationship between \a{ii} and latency.}
As expected, \a{ii} and latency have a monotonic relationship. However, changes in \a{ii} and latency are not proportional as the latter metric also captures architecture-specific characteristics. In particular, \a{ii} does not consider the different delay of instructions in the \a{isa}. In \textit{OpenEdgeCGRA}, load instructions from \autoref{tbl:instr_set} take 2 clock cycles, while other instructions require a single one. Therefore, because the \a{cgra}'s execution of an instruction is blocked by its slowest \a{pe}, packing all slow instructions in the same \a{cgra}-cycle might yield a faster execution than spreading them throughout the \a{cil}.
In addition, timing is affected when access to shared resources is serialized by the hardware. In our target architecture, concurrent store instructions to the same memory bank are pipelined. Furthermore, read instructions in the same column (even aiming at different memory banks) are also serialized. The arbitration among memory accesses hence increases run time, as the execution of \a{cgra} instructions must be paused until all memory accesses scheduled concurrently terminate. 
These effects can be observed in \autoref{tab:energy_latency} for \kstrs{} going from $D3$ to $D2$: the benchmarks can be mapped in the smaller architecture by doubling the \a{ii}, but such penalty is exacerbated by resources contentions, leading to a tripling of run-time latency. 

\begin{figure}[t]
\centering
\minipage[t][1cm][t]{0.75\textwidth}
  \includegraphics[width=\linewidth]{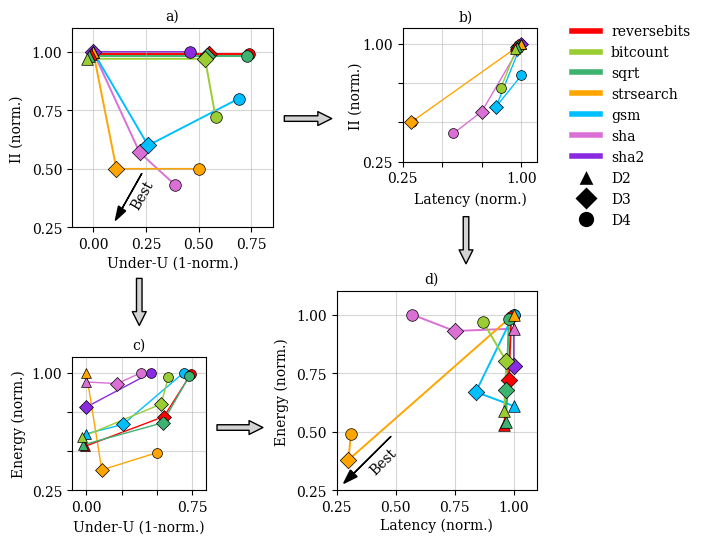}
\endminipage\hfill
\caption{a), d): Performance-efficiency solution space of mapped \as{cil}, in terms of compiler-level metrics~(a) and run-time metrics~(d). b),~c):~relation between performance and efficiency metrics in the two spaces. Data is reported for each \a{cil} across \a{cgra} sizes and is normalized per \a{cil}. Overlaps are slightly offset for visualization. 
}
\label{fig:energy_latency}

\end{figure}

\paragraph{Relationship between compiler-level and run-time metrics}
While the \a{ii} provides clues about  execution latency, similarly \a{u} (the ratio of non-idle \as{pe} across all mapped instructions)  gives hints about the power efficiency of a mapping. 
In this light, \autoref{fig:energy_latency}.a illustrates the performance/efficiency trade-offs exposed by different \a{cgra} sizes for the considered benchmarks using these compiler metrics\footnote{For clarity, the graph reports under-utilization (Under-U), that is, the ratio of idle \as{pe} over all mapped instructions, which is positively correlated to energy requirements.}.
\autoref{fig:energy_latency}.d instead reports the same trade-off space between performance and efficiency expressed with run-time metrics (energy and latency) from post-synthesis simulations. In all cases, values in \autoref{fig:energy_latency} are normalized per \a{cil} to the maximum among the three different sizes. \autoref{fig:energy_latency}.b and \ref{fig:energy_latency}.c plots the relation between corresponding metrics in the compiler-level and run-time space: \a{ii} vs. Latency,  Energy vs. Under-U, respectively. 
These intermediate plots highlight two trends. In \autoref{fig:energy_latency}.b, a decrease in \a{ii} caused by the size increase reflects in an almost linear drop in latency. In \autoref{fig:energy_latency}.c
the increase in U caused by the size decrease is reflected in a drop in energy consumption limited by a fixed cost (which is seen as a curved down-slope). 

However, there are outliers to these patterns. In particular, the \kgsm{} benchmark has a minimum latency for the $D3$ architecture, as mentioned above. Moreover,
the \kstrs{} overall energy consumption almost triples when going from $D3$ to $D2$ due to the drastic increase in latency with \a{ii}. This effect cannot be perceived in \autoref{fig:energy_latency}.a, but becomes evident in subplots b, c, and d, where implementation details are considered. 
These examples show that considering mapping results from a purely \a{ii} \cite{karunaratne2017hycube, mei2007adres, hamzeh2013regimap, dave2018ramp, friedman2019spr, pathseeker} or \a{u} \cite{wijerathne2021himap, mei2007adres} perspective does not completely capture the performance of mapped \as{cil}. Nonetheless, they can still be effectively employed to prune the design space of possible implementations, in advance of time-consuming detailed hardware simulations. 
In particular, the Pareto sets of best performing implementations (i.e., the ones which are not dominated in terms of performance/efficiency) is highly similar in \autoref{fig:energy_latency}.a and \autoref{fig:energy_latency}.d. Considering all benchmarks, 83\% of the Pareto points in the (\a{ii}, \a{u}) space are also Pareto points in the (energy,latency) space. Dually, all Pareto points in \autoref{fig:energy_latency}.d are part of the Pareto set in \autoref{fig:energy_latency}.d. Hence, only selecting Pareto solutions in the architecture-agnostic (\a{ii}, \a{u}) space for detailed (but lengthy and architecture-specific) hardware simulation does not result in sub-optimal outcomes, while pruning the exploration space by 40\%.

\section{Conclusion}
\label{sec:conclusion}

In this paper, we have presented a tool, called \toolname, for modulo scheduling \as{cil} onto \as{cgra}. We found that previous techniques, mainly based on classic graph algorithms such as Max-Clique enumeration, do not always explore the scheduling space effectively. Therefore, we have proposed a new \a{sat} formulation of the modulo scheduling problem on \a{cgra} that fully explores the scheduling space and finds the lowest \a{ii} possible for a given \a{dfg}.
To define the mapping problem through a \a{sat} formulation, we also introduce a new ad-hoc schedule called Kernel Mobility Schedule, which is used with the \a{dfg} of the \a{cil} to be mapped, and with the architectural information of the \a{cgra}, to generate all the constraints that the \a{sat} solver needs to obey. 
Overall, \toolname finds better solutions with respect to the \a{soa} alternatives\cite{dave2018ramp, pathseeker}, achieves better results in around 50\% of the cases, and even identifies valid mappings where other tools could not find a valid solution. Compiled \as{cil} where mapped on \textit{OpenEdgeCGRA} in ~\cite{OpenEdgeCGRA}. We show how the evaluation of architecture-agnostic compiler-level metrics can effectively guide the mapping process, even if it cannot provide a complete view of run-time efficiency and performance. Moreover, we demonstrate how it can be employed to successfully prune the design space in a fast way, in advance of laborious hardware explorations.

\begin{acks}
This research is carried out in the frame of the “UrbanTwin: An urban digital twin for climate action: Assessing policies and solutions for energy, water and infrastructure” project with the financial support of the ETH-Domain Joint Initiative program in the Strategic Area Energy, Climate and Sustainable Environment. This research was also partially supported by EC H2020 FVLLMONTI project (GA No. 101016776), and by the ACCESS – AI Chip Center for Emerging Smart Systems, sponsored by InnoHK funding, Hong Kong SAR. This work was also supported by the Swiss National Science Foundation under Grants: ML-Edge (200020-182009) and ADApprox (200020-188613).
\end{acks}


\bibliographystyle{ACM-Reference-Format}
\bibliography{main.bib}

\end{document}